\documentclass{emulateapj}
 
\newcommand{\epeak}{$E_{\rm peak}$}
\newcommand{\eb}{$E_{\rm b}$}
\newcommand{\kan}{{\it K06}}
\newcommand{\peakenergy}{${\mathcal E}_{\rm pk}$}

\begin{document}

\title{Broadband Spectral Properties of Bright High-Energy Gamma-Ray Bursts
         Observed with BATSE and EGRET}

\author{Yuki~Kaneko\altaffilmark{1}, 
         M.~Magdalena~Gonz\'alez\altaffilmark{2}, 
         Robert~D.~Preece\altaffilmark{3}, 
         Brenda~L.~Dingus\altaffilmark{4} and 
         Michael~S.~Briggs\altaffilmark{3}}

\altaffiltext{1}{Universities Space Research Association /
                        NSSTC, VP62, Huntsville, AL 35805,
                        yuki@sabanciuniv.edu}

\altaffiltext{2}{Instituto de Astronomia, 
                        Universidad Nacional Autonoma de Mexico, Mexico} 

\altaffiltext{3}{Department of Physics, University of Alabama in Huntsville /
                        NSSTC, VP62, Huntsville, AL 35805} 

\altaffiltext{4}{Los Alamos National Laboratory, Los Alamos, NM 87545}

\begin{abstract}
We present the spectral analysis of duration-integrated broadband spectra 
(in $\sim30\;$keV$-200\;$MeV) of 15 bright BATSE gamma-ray bursts (GRBs).  
Some GRB spectra are very hard, with their spectral peak 
energies being above the BATSE LAD passband limit of $\sim$2~MeV.
In such cases, their high-energy spectral parameters (peak energy and
high-energy power-law indices) cannot be adequately constrained by BATSE LAD 
data alone.
A few dozen bright BATSE GRBs were also observed with EGRET's calorimeter, TASC, 
in multi-MeV energy band, with a large effective area and fine energy resolution.
Combining the BATSE and TASC data, therefore, affords spectra that span four 
decades of energy ($30\;$keV$-200\;$MeV), allowing for a broadband spectral 
analysis with good statistics.
Studying such broadband high-energy spectra of GRB prompt emission is crucial,
as they provide key clues to understanding its gamma-ray emission mechanism.
Among the 15 GRB spectra, we found two cases with a significant high-energy 
excess, and another case with a extremely high peak energy 
(\epeak~$\gtrsim$ 170~MeV).
There have been very limited number of GRBs observed at MeV energies and above, 
and only a few instruments have been capable of observing GRBs in this energy 
band with such high sensitivity.
Thus, our analysis results presented here should also help predict GRB 
observations with current and future high-energy instruments such as AGILE and
GLAST, as well as with ground-based very-high-energy telescopes.
\end{abstract}


\section{Introduction}

High-energy observations of gamma-ray burst (GRB) prompt emission with 
various detectors have indicated that GRB continuum spectra extend up to 
MeV--GeV energies \citep{han94,hur94,cat98,din98,har98,kip98}.
GRB spectra in the MeV to GeV range are usually well-described by a single 
power-law with an index in the approximate range of $-1$ to $-4$
\citep{sch92, kwo93, hur94, han94, sch95, cat98, kip98, bri+99, wre02}.
This range is in agreement with the distributions of the high-energy 
power-law indices observed with BATSE Large Area Detectors (LADs) in
$\sim$30~keV$-$2~MeV \citep[\kan~hereafter]{kan06}.
However, due to the power-law nature of GRB spectra, photon counts above 
$\sim$1~MeV are usually very low, and this, combined with the fact that the 
field of view of a high-energy detector is generally limited, results in
much fewer GRBs observed in multi-MeV band than keV-band observations.

The Burst and Transient Source Experiment (BATSE) on board the {\it Compton
Gamma-Ray Observatory (CGRO)} has provided the largest GRB spectral database to 
date in the passband of sub-MeV range.  
Previously, \kan~showed that there are some BATSE GRB spectra whose 
peak energy (\epeak) lies very close to or above the upper energy bound 
($\sim$2~MeV) of BATSE LADs.
For such cases, the LAD data alone cannot adequately determine either \epeak~or
the high-energy power law index.
Moreover, the ability to identify the high-energy power law component using LAD data
alone is limited, since the LAD sensitivity decreases significantly toward the 
upper energy bound \citep{fis89}. 
Therefore, observations of spectra extending to much higher energies with 
reasonable sensitivity are needed for these spectral parameters to be well 
determined.
Combining BATSE LAD data with multi-MeV observations by another high-energy 
detector enables such a broadband study.

Broadband spectral analyses of a few BATSE GRBs have been presented in the 
literature, using the data obtained with the \textit{CGRO} instruments 
\citep{sch98, bri+99}.
However, those broadband spectra were superpositions of the deconvolved photon 
spectra that were obtained by analyzing each dataset separately with various 
photon models.
Deconvolved photon counts are model dependent, and a spectrum constructed by
combining individually deconvolved spectra can be quite different from that
obtained properly by simultaneously fitting a common model to all datasets.

The Total Absorption Shower Counter (TASC), the calorimeter of another
experiment also aboard the {\it CGRO} -- the Energetic Gamma-Ray Experiment 
Telescope (EGRET) spark chamber -- was one of the few instruments 
capable of observing GRBs in the multi-MeV energy band with excellent energy 
resolution \citep{tho93}.
Due to its sensitivity to all directions, some bright, BATSE-triggered GRBs 
were also observed with the TASC apart from the EGRET spark chamber events.
Since the TASC provided spectra in the range $\sim$1$-$200 MeV, broadband GRB 
spectra spanning four decades of energy can be obtained by combining LAD and 
TASC data. 
Such spectra, in at least one case, resulted in the discovery of a distinct 
multi-MeV spectral component apart from the extrapolated sub-MeV BATSE component 
\citep{gon03}.  
In this paper, we present duration-integrated broadband spectral analysis 
of 15 bright BATSE GRBs, using LAD and TASC data.
We first describe the instuments and available data types of TASC in 
\S\ref{sec:instruments}, and the selection and analysis methodology of our study
in \S\ref{sec:tasc_selection} \& \S\ref{sec:tasc_analysis}.  
Then in \S\ref{sec:tasc_result}, we present 
the results, and finally discuss the results in \S\ref{sec:sum}.

\section{Total Absorption Shower Counter (TASC)}\label{sec:instruments}
EGRET spark chamber was designed to observe high-energy gamma rays much above 
the BATSE energy band, between $\sim$20~MeV$-$30~GeV.  
It was equipped with an anti-coincidence counter and a calorimeter, TASC, 
which was located at the bottom of the module.
Although the field of view of the EGRET spark chamber was limited
\citep{din98}, the TASC was capable of accumulating data for BATSE-triggered 
GRBs from all directions, independently from the spark chamber events.

\begin{figure}
\epsscale{0.8}
\centerline{
\plotone{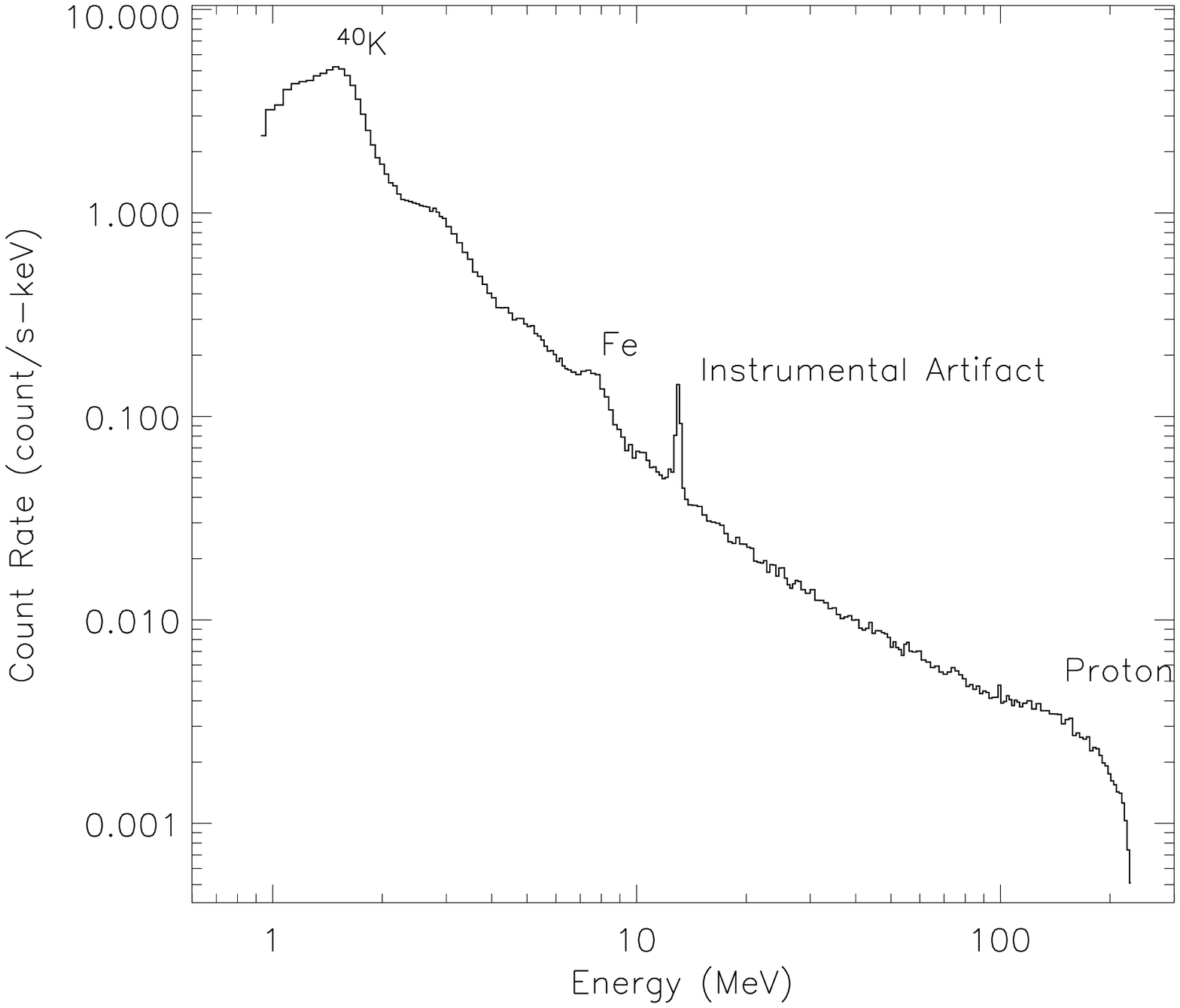}}
\caption{An example TASC raw count spectrum of GRB~910503, for the time
interval including the burst episode ($-9.8$ to 121$\;$s since BATSE trigger time).
The count rate includes background and the source counts.  $^{40}$K line, 
Fe line, the instrumental artifact, and the cosmic-ray proton features are 
clearly seen.}
\label{fig:tasc_sp}
\end{figure}

Like the BATSE detectors, the TASC was also a NaI(Tl) scintillation detector but
with much larger dimensions of $76\;$cm $\times~76\;$cm with a 20-cm thickness 
(corresponding to 8$\;$radiation lengths).
In its low-energy mode in the energy range of $\sim$1$-$200$\;$MeV, the TASC 
continuously accumulated the non-burst Solar spectra (SOLAR), every 32.768$\;$s.
In addition, it collected Burst spectra (BURST) initiated by BATSE triggers, 
in four commandable time intervals of 1, 2, 4, and 16 (or 32)$\;$s.
Both the SOLAR and BURST data provided spectra with 256 energy channels.
An example of TASC raw count-rate spectra (i.e., including background counts) 
of GRB~910503 integrated over the burst duration is shown in 
Figure~\ref{fig:tasc_sp}.  
In the spectra, the $^{40}$K line at 1.46$\;$MeV and the Fe neutron capture 
line at 7.64$\;$MeV are always present, and they were used for on-board 
calibration purposes.  
The instrumental artifact at $\sim$12$\;$MeV is due to an error in the 
electronics design \citep{tho93}.  
There is also a bump around 100$-$200$\;$MeV caused by cosmic-ray protons that 
passed through the TASC along the z-axis and deposited energy of $\sim 105\;$MeV.
The proton spectral feature was also used to monitor the gain of the TASC.
The FWHM energy resolution of the TASC is about 20\% over its entire energy 
range.

The response is highly dependent on the incident direction of the event photons,
because of the block shape of the TASC NaI crystal, as well as the presence of
intervening spacecraft materials surrounding the detector. 
The TASC was not capable of localizing events; therefore, for GRB observations,
the locations determined by BATSE were used to obtain detector response for
each event.  The response was calculated using EGS4 Monte Carlo code 
\citep{nel85} with the complete {\it CGRO} mass model.
It should be noted that the deadtime of the TASC is extremely high, $\sim$60\%
on average.

\section{Selection Methodology}\label{sec:tasc_selection}

We searched the TASC data for the trigger times of 43 bright BATSE GRBs.
The 43 GRBs were selected based on the criteria of peak photon flux 
$\geq 10\;$photons$\;$cm$^{-2}\;$s$^{-1}$ in the 1024$\;$ms timescale and BATSE 
channel-4 energy fluence ($>$ 300 keV) 
$\geq 5.0\;\times\;10^{-5}\;$ergs$\;$cm$^{-2}$. The flux and fluence values were
taken from the 4B \citep{pac99} and the current\footnotemark{} BATSE catalog.
Among these, we identified 28 GRBs to have significant detections in the 
TASC data.  
In this analysis our sample consists of 15 bursts (out of 28), for which we had 
computed TASC detector response matrices prior to this study.
The 15 GRBs are listed in Table~\ref{tab:joint_events} along with the time and 
energy intervals used for analysis, for which the selection methods are 
described below.
\footnotetext{See http://gammaray.nsstc.nasa.gov/batse/grb/catalog/current/.} 
 
For this analysis we used TASC SOLAR data, which provides 32.768-s time 
resolution.
Since all of the 15 events are very strong and most of them have durations longer than 
33 seconds, the time resolution is adequate for the duration-integrated spectral 
analysis performed here.
As for the selection of corresponding BATSE data, we use the LAD Continuous 
(CONT) data of the brightest LAD, with 2.048-s time resolution.
The brightest LAD (i.e., the LAD that recorded the highest counts) for each 
event is also listed in Table~\ref{tab:joint_events}.
The CONT data were able to provide time intervals with a sufficient match to 
the SOLAR data time intervals, which often began before the BATSE trigger time.
The time intervals chosen for analysis were determined based upon the 
detection significance above background for either the LAD or TASC lightcurves.
For all but one event (with a very weak tail in the LAD data), the time 
intervals chosen for the analysis contained their BATSE T$_{90}$ intervals.
All 15 events are also included in the BATSE LAD spectral analysis presented in
\kan, although the data type used here (hence the time and
the energy intervals) may be different.
In Figure~\ref{fig:lad_tasc_lc}, we show, for all 15 GRBs, both the LAD 
and TASC lightcurves over their entire energy ranges.  
In the Figure, the background models and the time intervals used for the 
analysis are also shown.
The background models were determined by fitting a low-order polynomial function
to the spectra accumulated for several hundreds of seconds before and after the 
burst duration interval.

LAD CONT data provided 16 energy channels in the energy range of 
$\sim$30$\;$keV$-$2$\;$MeV.  The lowest few channels are usually below the 
electronic threshold, and the highest channel is an energy overflow channel;
therefore, a total of about 13 energy channels were usually included in the 
analysis.
For the TASC SOLAR data, the lowest 6$-$7 channels are always excluded to 
assure the exclusion of an electronic cutoff.  
This translates into a lower energy bound of $\sim$1.3$\;$MeV (Table 
\ref{tab:joint_events}, column 9).
In addition, the uppermost 10$-$20 channels are also excluded, depending upon 
the gain of the detector at the time of the event.  
The resulting upper bound for the TASC energy range was $\sim$130$-$200$\;$MeV
(Table \ref{tab:joint_events}, column 10).
Notice that overlap in energy between the LAD and TASC datasets of a few 
hundred keV exists in each event.

\section{Spectral Analysis}\label{sec:tasc_analysis}
We converted the TASC data into the BATSE BFITS format, and used the BATSE 
spectral analysis tool RMFIT\footnotemark{}
\footnotetext{R.S.~Mallozzi, R.D.~Preece, \& M.S.~Briggs, "RMFIT, A Lightcurve
and Spectral Analysis Tool," \copyright~Robert~D.~Preece, University of Alabama
in Huntsville}for this analysis.
For comparison purpose and consistency, the same set of photon models used to 
analyze the bright BATSE GRBs in \kan, namely, a single power-law 
(PWRL), empirical GRB functions with and without the high-energy power-law 
(BAND and COMP, respectively; \citealt{ban93}), and the smoothly-broken 
power-law (SBPL; \citealt{wingspan, ryd99, pre00}) with its subsets, were fitted to the joint LAD--TASC dataset.
Extensive discussions on these spectral models are found in \kan.~
Since we are only analyzing the duration-integrated spectra, the GRB 
function with fixed high-energy index $\beta$ (BETA model in \kan) was not 
applicable and thus not used.
The PWRL and COMP models are expected to result in very poor fits to the joint 
spectra because of the broad energy coverage, as well as the fact that our 
sample contains the very brightest of all BATSE GRBs, which have been shown to 
have a smoothly-broken power-law behavior with no cutoff in keV band (their 
best-fit models in the LAD-only analysis are SBPL or BAND; \kan). 

In order to account for uncertainties in the effective areas of the two 
detectors, we added an effective area correction (EAC) term to the spectral model
of each fit: this is a multiplicative factor, to normalize the data of the 
second detector to those of the first detector.
In our case, the EAC factor always normalized the TASC data to the LAD data.
The EAC values can vary from event to event and were determined by simultaneously 
fitting the BAND model to both datasets with the EAC term as a free parameter.
Once the EAC factor was found, it was kept fixed in the subsequent final fits.
Between the LAD and the TASC datasets, they were always found to be unusually 
large ($\sim$0.5).
We investigate this issue further in \S \ref{sec:eac}.

\section{Analysis Results}\label{sec:tasc_result}
As we predicted, 
broadband high-statistics data confidently rejected PWRL and COMP as best-fit
models for all 15 spectra in our sample.
Following the LAD GRB analysis in \kan, we determined the best-fit (BEST) models 
for the 15 joint spectra as well.
The BEST model is the simplest, statistically well-fit model among the four
spectral models described above that is required by the observed spectrum.
Since we have four models with 2, 3, 4, and 5 free parameters, we compared 
resulting $\chi^2$ values between all combination of models, and searched for
significant ($>$99.9\%) improvement in $\chi^2$ as we move from the simplest 
to more complicated models.  We also take into account the spectral parameter
constraints. The details of determining the BEST models is found in \kan.

For all GRBs, the joint count spectra and the BEST models, along with the 
corresponding $\nu F_{\nu}$ spectra are shown in Figure~\ref{fig:lad_tasc_cntsp}.
The spectral parameters of the BEST model are also presented in 
Table~\ref{tab:joint_results}.
In the table, \peakenergy~value\footnotemark{} is the peak energy in 
$\nu F_{\nu}$ spectrum and break energy (\eb) value is the break energy of 
a broken power law, regardless of the model fitted. 
\footnotetext{Hereafter, we use \peakenergy~notation to represent the actual 
peak energy in $\nu F_{\nu}$ spectrum, to distinguish from a fit parameter 
\epeak~in BAND model. \epeak~$=$~\peakenergy~only if $\beta < -2$.}
As mentioned in \kan, \epeak~and \eb~are not necessarily 
the same for a given spectrum due to the curvature around the break energy.
The derivation of these values are presented in appendices of \kan.
It must be noted that due to the high photon counts of these events, 
uncertainties in the data are dominated by systematics, which can be large
especially at lower energies ($\lesssim 100\;$keV) in the LAD data.
This can result in a relatively large $\chi^2$ value even for an acceptable fit,
which can be seen in the sigma residuals in Figure~\ref{fig:lad_tasc_cntsp}.

\begin{figure}
\plotone{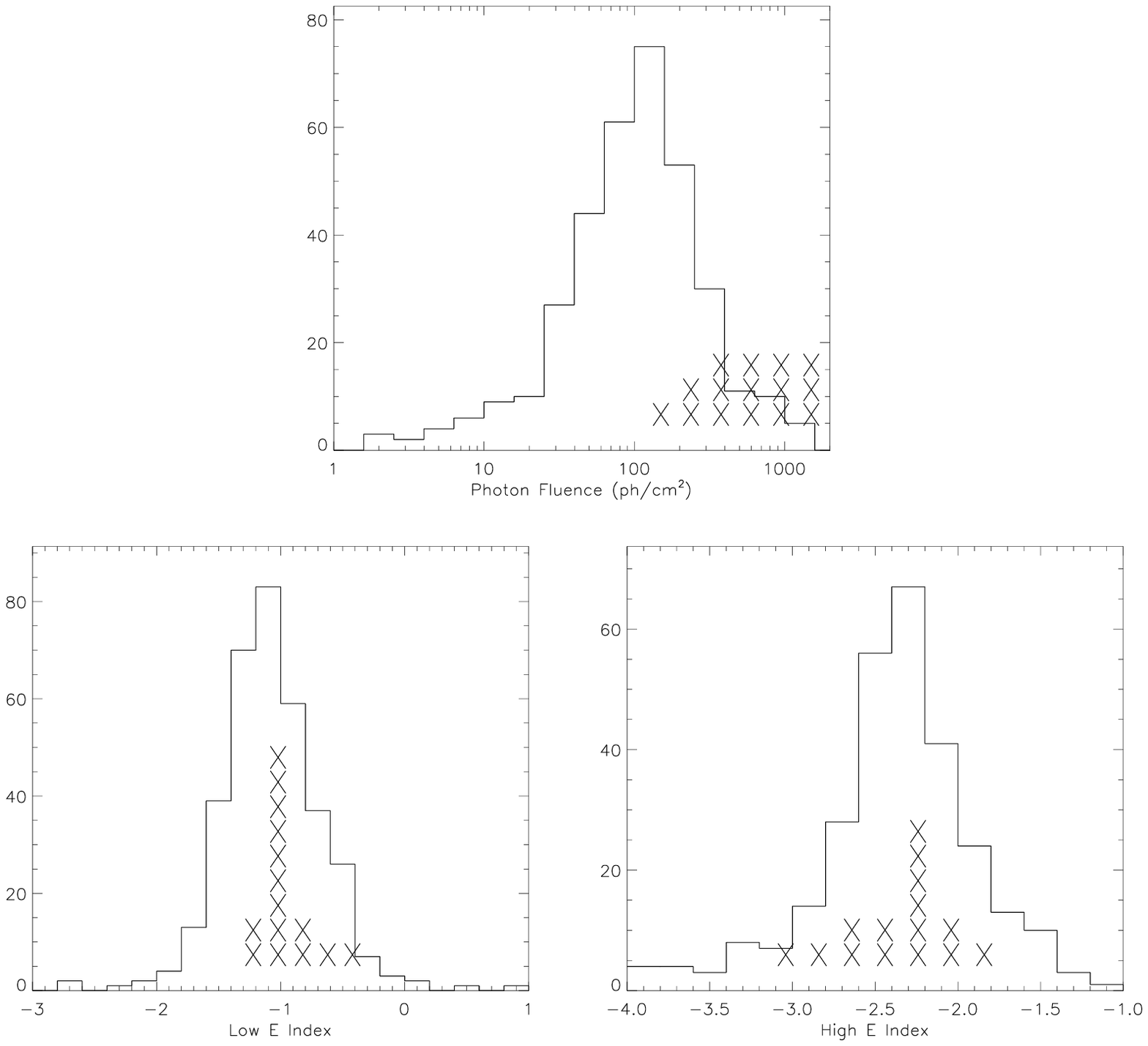}
\caption{Parameter distribution comparisons of 15 joint events
with 342 bright BATSE GRBs.
The crosses represent photon fluence ({\it top}) and spectral indices 
({\it bottom}) of jointly analyzed events, and histogram shows the distribution 
of LAD-only analysis events.}
\label{fig:joint_hist1}
\end{figure}

\begin{figure}
\plotone{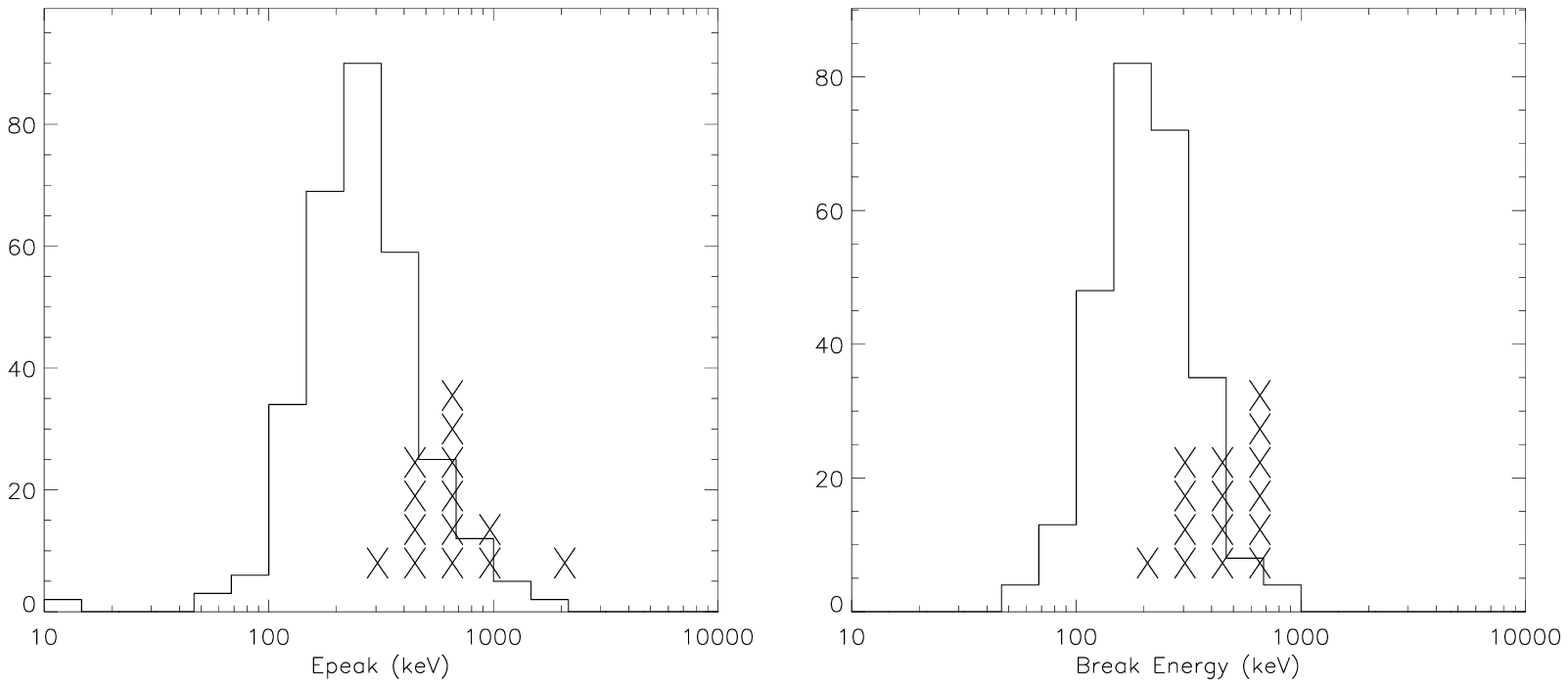}
\caption{Parameter distribution comparisons of 15 joint events with 
342 bright BATSE GRBs.
The crosses represent \peakenergy~({\it left}) and \eb~({\it right}) 
values of jointly analyzed events, and histogram shows the distribution of 
LAD events.  In the \peakenergy~distribution, GRB~930506 (trigger 2329) is 
excluded due to a very high \peakenergy~$>167$~MeV.}
\label{fig:joint_hist2}
\end{figure}

To compare these parameters with the parameter distributions of the larger 
sample of bright BATSE bursts presented in \kan, we overplot in 
Figures~\ref{fig:joint_hist1} and \ref{fig:joint_hist2} the BEST model 
parameters of the jointly-analyzed events on top of the time-integrated LAD 
spectral parameter distributions, taken from \kan.
It is evident in terms of the photon fluence (in 25$-$2500~keV; 
Figure~\ref{fig:joint_hist1}, top panel) that these 15 events are in the very 
brightest group of all BATSE GRBs. No bias or tendency is seen in the 
distributions of spectral indices (Figure~\ref{fig:joint_hist1}, bottom panels).
It is clear, however, that the \peakenergy~for the 
15 events belong to the higher end of the BATSE distributions 
(Figure \ref{fig:joint_hist2}), and by themselves forms a quasi-Gaussian
with a median of $\langle$\peakenergy$\rangle = 517$~keV.
A likely reason for this is because our sample selection was based on the high 
photon flux and fluence above 300 keV: a group correlation between burst 
brightness and \peakenergy~was previously identified using an early BATSE GRB 
sample by \citet{mal95}.
We note also that because we selected only GRBs with significant 
detections in TASC data ($\gtrsim$~1~MeV), this naturally introduces a 
preference for GRBs with higher \peakenergy~to be included in our sample.
We found that the photon fluence and the energy fluence determined with the 
joint spectra were consistent with the values estimated by extrapolating the 
LAD-only fit spectra up to 200~MeV.
In addition, we did not find significant correlation between photon fluence 
and \epeak~or \eb~within our sample.

\begin{figure}
\epsscale{1.1}
\plottwo{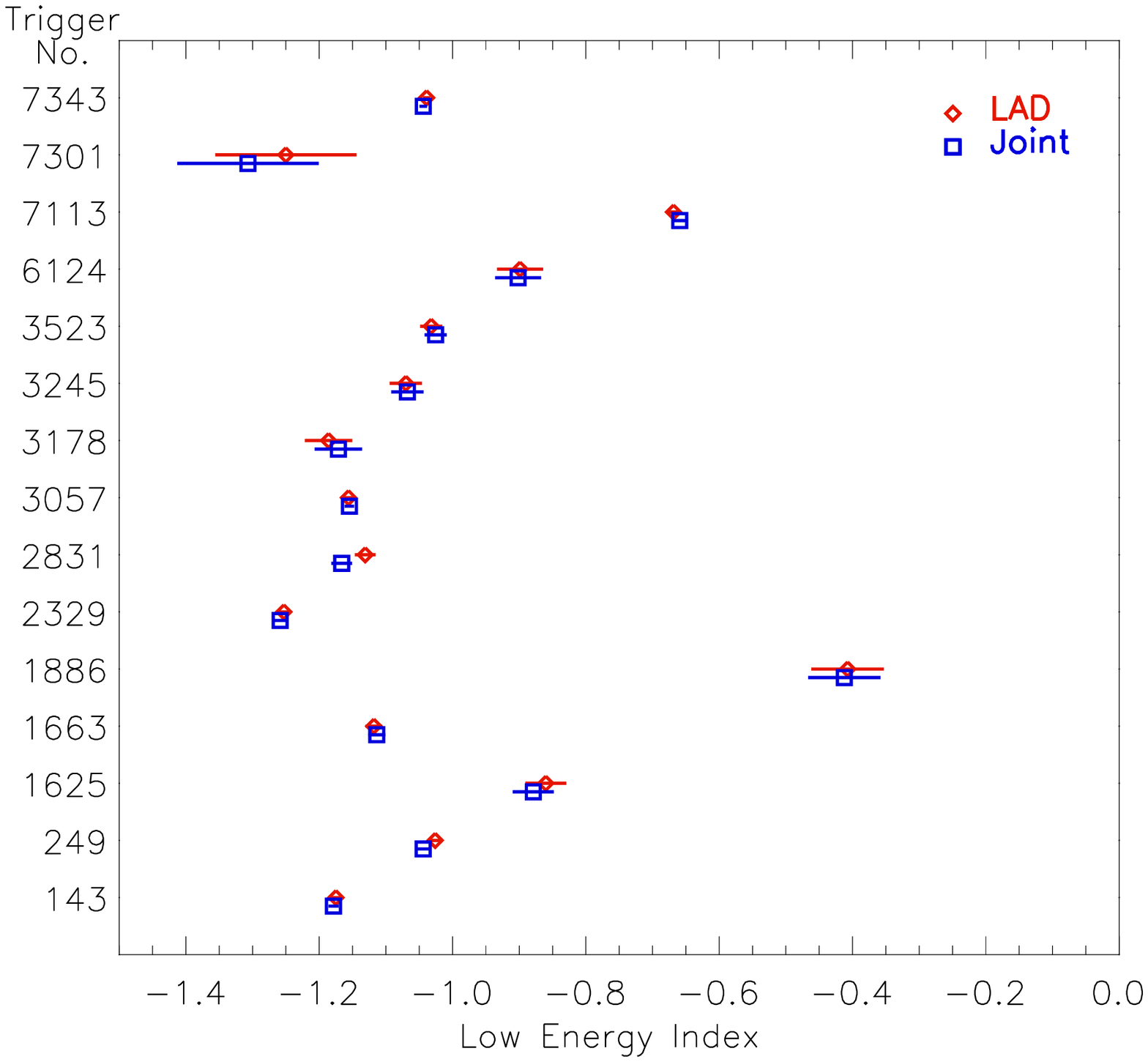}{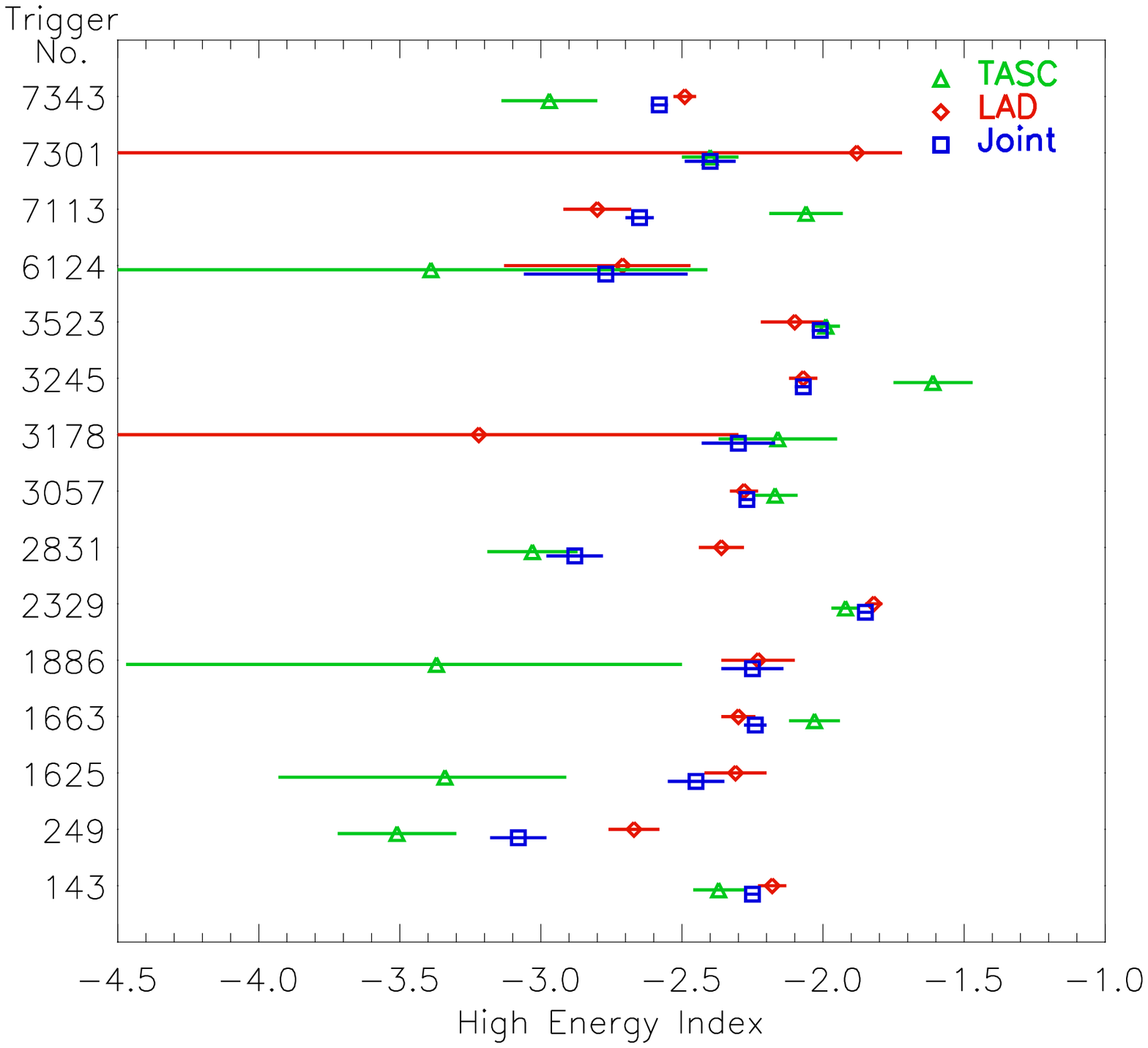} \\
\plottwo{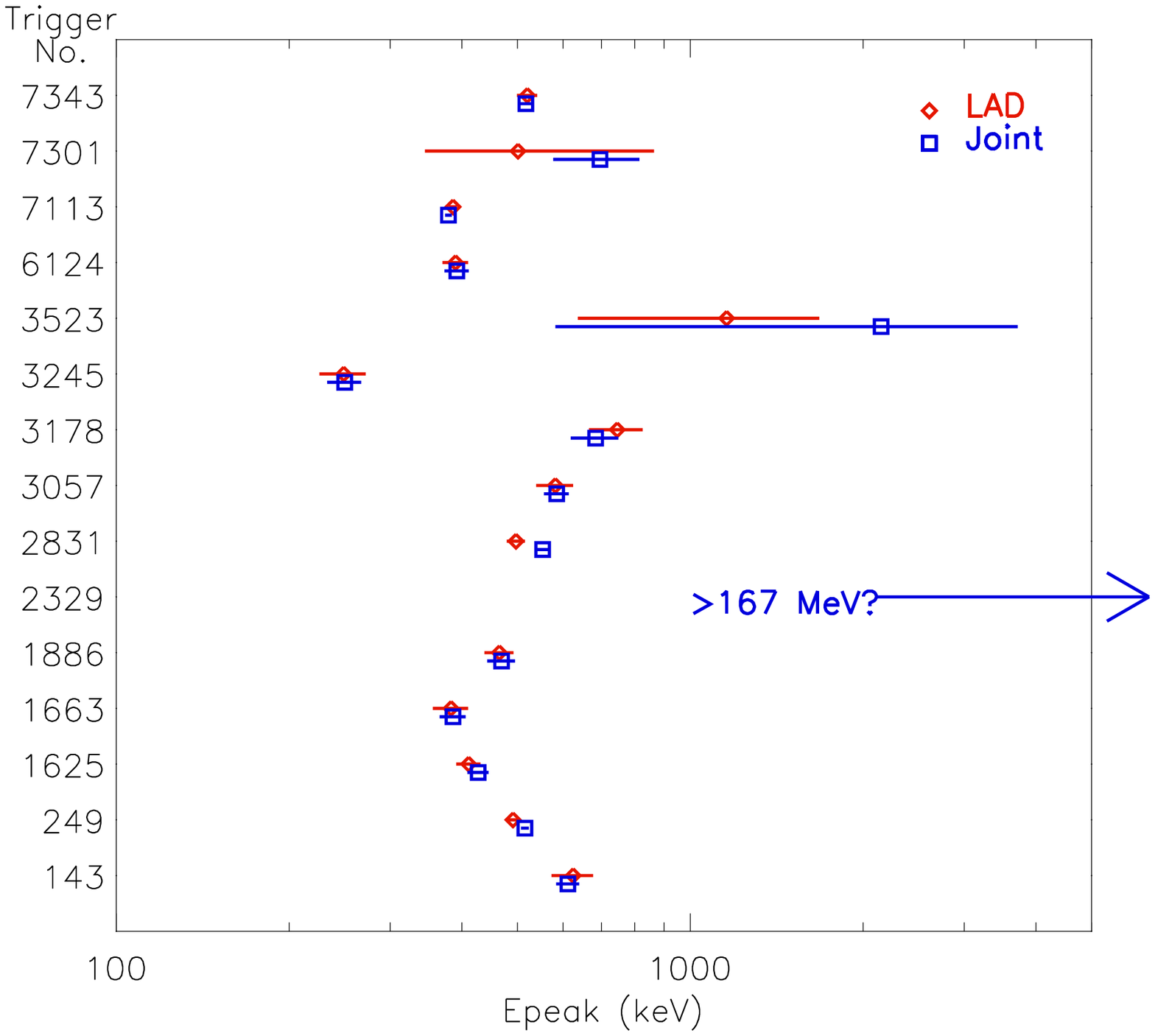}{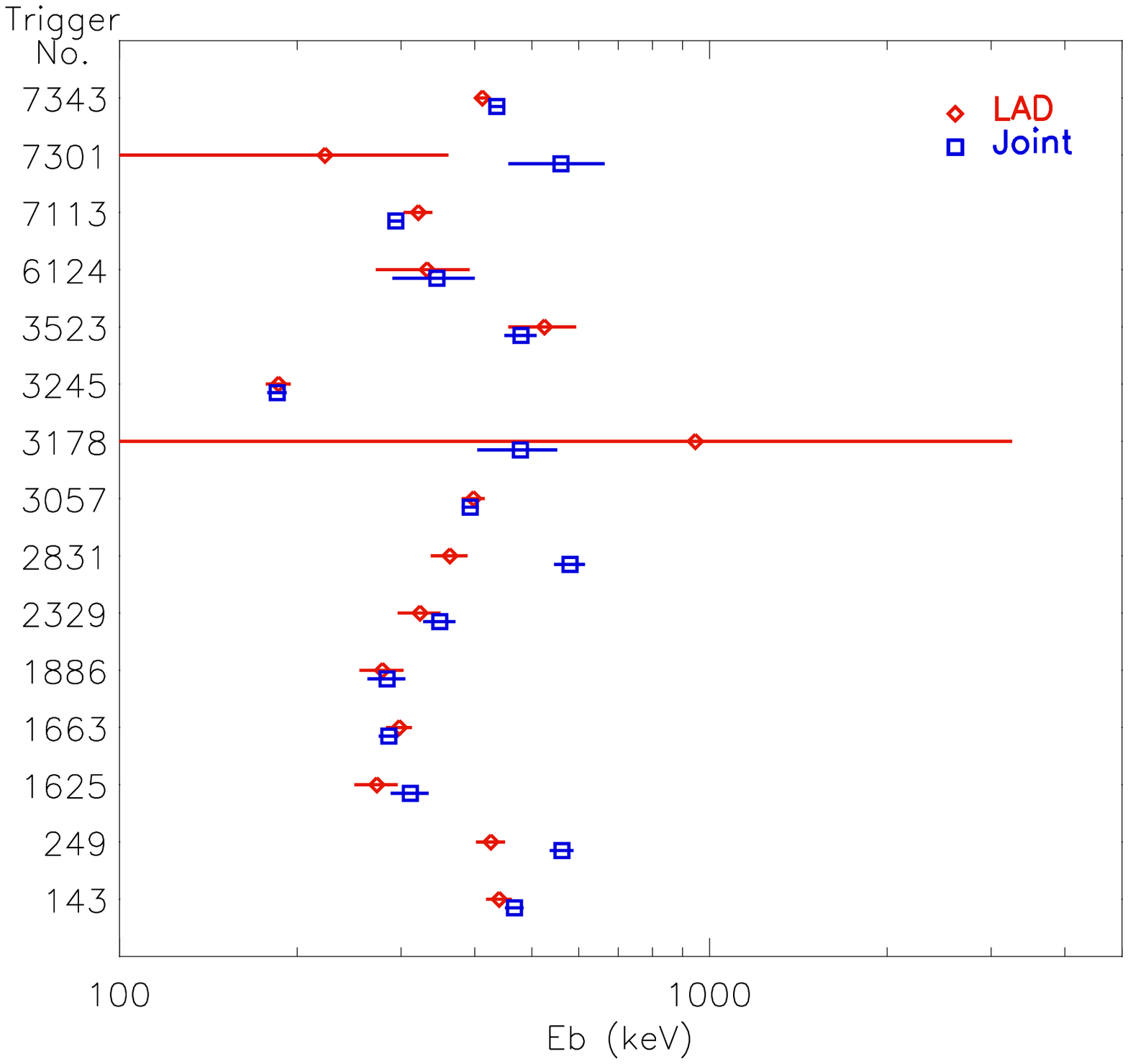}
\caption{Low- \& high-energy indices ({\it top panels}) and \peakenergy~and 
\eb~({\it bottom panels}) determined by joint analysis (squares) 
and individual analysis of LAD (diamonds) and TASC (triangles).
Left axis shows the event trigger numbers. All uncertainties are 1$\sigma$.
\peakenergy~value of trigger 2329 is $> 167$~MeV.}
\label{fig:joint_par}
\end{figure}

To illustrate the improvements in parameter constraints as a result of the joint 
analysis, in comparison with the single-detector analysis, we show the spectral 
parameters determined by the joint fits and the individual detector fits
in Figure~\ref{fig:joint_par}.
In the single-detector analysis, the LAD data were fitted with the BEST model 
listed in Table \ref{tab:joint_results} while the TASC data were fitted with
PWRL with pivot energy of 10~MeV. 
The PWRL indices of TASC are compared with the high-energy indices of the BEST 
models in Figure~\ref{fig:joint_par} (top right panel).
It is not surprising that in most cases the spectral parameters are determined 
much better with the joint analysis than with the individual cases. 
We also note that spectral parameters of our TASC-only fits were consistent 
with the previous TASC analysis results found in the literature 
\citep[e.g., ][]{sch92, kwo93, hur94, cat98, bri+99, wre02}
within a few sigma uncertainties. 
Moreover, a few of the 15 GRBs were also detected by the EGRET spark chamber 
in even higher energies, with their flux values consistent with the TASC spectra 
\citep{sch92, kwo93, hur94}.

As seen in Figure~\ref{fig:joint_par} (top left panel), the low-energy spectral
indices are already well constrained by the LAD data alone and are least 
affected by the addition of TASC data.
On the other hand, the high-energy indices are much better constrained by the 
joint analysis, as one would expect.  
The values determined by the joint analysis are nearly always found between the
values derived from the LAD-only and TASC-only analyses.
In cases where the LAD-determined index differs from the jointly-determined 
index by more than a few $\sigma$ (i.e., triggers 249 and 2831),  
the spectral break energy was also found to change significantly 
(Figure~\ref{fig:joint_par}, bottom right panel).
The \peakenergy~values found by the joint analysis are consistent with those
found by the LAD-only fits, although in many cases it seems to settle in the 
higher ends of the value ranges determined by LAD.  
We observed one case, trigger number 3523, in which the value of 
\peakenergy~determined with the joint fit was less constrained than the one 
found with the LAD-only fit.
This is due to the high-energy index being very close to $-2$ ($-2.01 \pm 0.03$), 
making the constraint of \peakenergy~very difficult, by definition (see its 
$\nu F_{\nu}$ spectrum in Figure~\ref{fig:lad_tasc_cntsp}).
There was one event (trigger number 2329) for which the \peakenergy~could not be 
determined even with the joint spectra because the high-energy index was above 
$-2$ by 7.5$\sigma$ (see Table~\ref{tab:joint_results})
and therefore, the $\nu F_{\nu}$ spectrum did not peak within the energy range.
This may indicate a lower limit in \peakenergy~of 167~MeV (upper energy bound of
the spectrum) for this event.
The lightcurve of this burst (Figure~\ref{fig:lad_tasc_lc}) shows that in the
first time interval (T--33 to 0~s), the TASC flux is clearly dominant and thus,
spectrally harder than the second interval spectrum.
We note that the time-resolved LAD-TASC joint analysis of this event in a 
time interval of 1$-$23 seconds also found that the high-energy indices were 
always above $-2$ by at least 1$\sigma$ \citep{gon04, kan05}.

Although the spectral fits presented in Table~\ref{tab:joint_results} were all
sufficiently good (large $\chi^2$ were due to systematics in LAD), indications 
of a high-energy excess were found in the residual patterns of at least three 
spectra (GRBs~920902, 941017, and 980923; triggers 1886, 3245, and 7113).
This was much more evident in the spectrum of 3245: The high-energy excess above 
$10\;$MeV is seen in sigma residuals of the spectrum 
(Figure~\ref{fig:lad_tasc_cntsp}), and this is also evident from its lightcurves 
(Figure~\ref{fig:lad_tasc_lc}).
For this event, fitting an additional high-energy PWRL together with the BEST 
model resulted in an improvement in $\chi^2$ of 18.1 (for $\Delta {\rm dof} = 2$), 
corresponding to a chance probability of $10^{-4}$.  
In the case of 7113, the $\chi^2$ improvement for adding high-energy PWRL was 
14.0, indicating a chance probability of $\sim$$10^{-3}$, whereas for the last
event (1886), the corresponding $\chi^2$ improvement was only about 7, which 
translates into a chance probability of 0.03.
Although, in all case, the additional high-energy PWRL indices could not be 
well determined (1$\sigma$ uncertainties $>$ 0.5), the indices were rather
hard, likely $> -2$.
Time-resolved analysis of 3245 previously revealed, with much higher 
significance, a high-energy spectral component that deviates from the 
extrapolated keV LAD component \citep{gon03}.
This high-energy component emerged later than and remained bright  
longer than the keV spectral component.
Although our analysis presented here is only for duration-integrated spectra,
the delayed nature of the excess high-energy emission is clear from the 
lightcurves of all three events (Figure~\ref{fig:lad_tasc_lc}).

\subsection{Effective Area Correction Issue}\label{sec:eac}
All detector response models have uncertainties.
To allow for the uncertainties in their effective areas, we employ
normalization factors between detectors when simultaneously 
analyzing spectra from multiple detectors.
Usually only about 10\% difference between datasets is expected \citep{bri+99}.
In our joint analysis, however, we find the discrepancy between the LAD and 
TASC data to be relatively large, $\sim$30$-$80\%, as indicated by the EAC 
factors (see Table \ref{tab:joint_results}).
Disagreements between LAD and TASC data were also previously found in some of 
the composed spectra \citep{sch98, bri+99}, although the response matrices used 
here have been newly calculated.

\begin{figure}
\epsscale{0.8}
\centerline{
\plotone{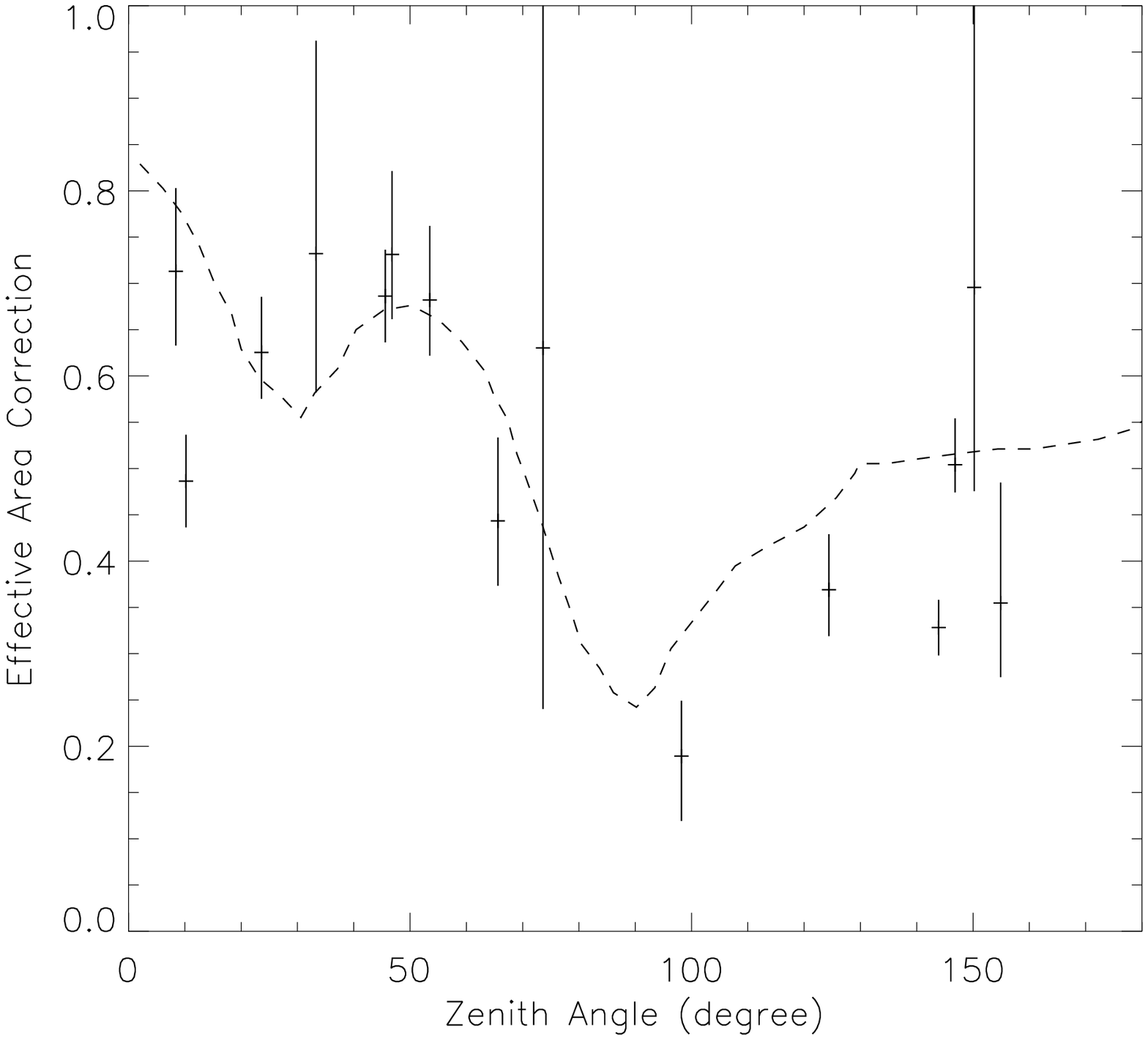}}
\caption{Effective area correction values of 15 events as a function of their
incident zenith angles. Dashed line shows the normalized TASC effective area 
\citep[taken from][]{tho93}, also as a function of incident zenith angle, and for
azimuth angle of 168$^o$. Uncertainties in EAC values are 1$\sigma$.}
\label{fig:tasc_norm}
\end{figure}

As mentioned earlier, the TASC effective area depends highly upon the incident 
angles of events, because of the geometric area of the NaI crystal as well as 
varying amount of intervening spacecraft material \citep{tho93}.
Therefore, we investigated the EAC factors in our sample in terms of the 
incident angles, as well as other potential contributing factors, such as the 
TASC live time, event brightness, energy range, and spectral parameters.
We found no apparent correlation in any of those; however, we noticed a striking 
resemblance between the plots of EAC vs.~incident zenith angle and of the 
effective area vs.~zenith angle.  The comparison is shown in 
Figure~\ref{fig:tasc_norm}, in which our EAC values for the 15 GRBs and the 
normalized TASC effective area are plotted against incident zenith angles.
This indicates that the discrepancy between the two detectors is more severe 
at the zenith angle where the TASC effective area is smaller.
As a matter of fact, a disagreement between the calculated effective area and 
the actual experimental value was found at the time of the TASC instrument 
calibration, 
which was attributed to the \textit{CGRO} mass model underestimating the 
intervening material \citep{tho93}.
The EAC factors in our analysis were always found to be less than 1, meaning 
count rates in the TASC data are overestimated.
The count overestimation becomes more apparent when there is larger amount of
intervening material, namely, when the effective area is smaller.
Consequently, we conclude that the \textit{CGRO} mass model used to determine the
effective area indeed underestimates the intervening material, at each zenith
angle, resulting in overestimation of the photon counts observed with TASC.

\section{Summary and Discussion}\label{sec:sum}
In this study, we extended the BATSE GRB spectral analysis (\kan) to high-energy 
broadband spectra obtained by combining BATSE LAD data and EGRET TASC data.  
Time-integrated joint spectra of 15 hard BATSE GRBs were analyzed in order to 
probe high-energy spectral properties of prompt emission.
The TASC data in multi-MeV energy band with fine energy resolution, clearly 
confirmed that the GRB spectra do extend up to $\sim$200~MeV and probably beyond.
The joint broadband spectra in the energy range of $\sim$30~keV$-$200~MeV indeed 
constrain the high-energy spectral indices and break energies of strong GRBs that 
have significant MeV emission much better than single detector analysis.
In most cases, \peakenergy~(and \eb) values derived with LAD data
alone and values found by joint analysis were consistent within 1$\sigma$
uncertainty.
However, in a few cases, the jointly-fit indices differed significantly from 
those determined with the LAD data alone.
This indicates the possibility that some high-energy indices obtained with 
BATSE LADs alone may not reflect the intrinsic values, and thus highlights the
importance of broadband spectral analysis.

We identified one case (GRB~930506; trigger 2329) in which the \peakenergy~could 
be extremely high, with a lower limit of $\sim$167 MeV.
We note here, however, that this possibly very high \peakenergy~should be
interpreted with caution for two reasons: 
\begin{enumerate}
\item{While the high-energy power law index being $> -2$ is statistically
very significant (by 7.5$\sigma$) as determined by the {\it statistical} error 
bar, there is the possibility of a systematic error existing in a joint fit 
between two instruments, which is not taken into account here.
Including the systematic uncertainties in the analysis could bring down the 
index value to $< -2$, in which case the 
\peakenergy~would probably become a few MeV.}
\item{Another possibility is that there are two spectral components in this
spectrum; the ``regular" one with \peakenergy~of a few MeV and an additional
high-energy multi-MeV component.
In addition, if the spectra evolved similar to the case of GRB~941017, the
duration-integrated spectrum could be a superposition of two evolving components. 
In such cases, the single model used here could hinder correct identification
of the \peakenergy~as well as the high-energy power law index.
We found, however, no statistical evidence for such high-energy component 
in the spectrum of GRB~930506.  
}
\end{enumerate}
In any case, the \peakenergy~of GRB~930506 is clearly above the BATSE LAD
energy range.
This, combined with the fact that in some cases the \peakenergy~values were
found to be slightly higher than those determined by LAD data only, may indicate 
that there exists a tail population of \peakenergy~values extending to a few 
hundred MeV, especially given the limited sensitivity of TASC.

In addition, while the broadband spectra were mostly consistent with broken 
power laws (BAND or SBPL), indications of high-energy excess were also 
found in two events (GRBs~941017 \& 980923; triggers 3245 \& 7113).
For one of them (GRB~941017; trigger 3245), the distinct MeV component has been 
clearly identified with time-resolved spectral analysis \citep{gon03}, 
with much stronger significance than with the time-integrated spectrum.
For this particular GRB, the existence of the extra component was further
reinforced by the COMPTEL spectra, which covered an energy range of
30\,keV$-$10\,MeV \citep{kan04}.

Possible explanations that have been proposed for such a high-energy component 
include the Compton upscattering of synchrotron photons in the reverse shock
by the synchrotron-emitting relativistic electrons \citep{gra03}, and the 
electromagnetic cascade emission of ultra-relativistic baryons through
photo-pion interactions and subsequent pion decay \citep{der04}.
In the synchrotron shock model \citep{kat94}, the synchrotron self-Compton emission 
associated with the synchrotron component that peaks in the keV band is expected 
to be observed in the MeV to TeV energy band, depending on the Lorentz factor of the
electrons \citep{gue03}. 
However, the brightness and delayed nature of the high-energy component 
are inconsistent with the synchrotron-self Compton.
It is particularly interesting if the component is indeed due to the 
relativistic baryons, since the observed component may be direct evidence of
baryonic acceleration, namely, cosmic rays.

The high-energy power-law component observed in GRB~941017 by \citet{gon03}
was described with an additional power law of index $\sim -1$.
This requires that there exists another break energy (and a true \peakenergy) 
above 200~MeV, in order to avoid the energy divergence.
Although the redshift for this event is unknown, a rough upper limit energy for
such a break could be placed at $\sim$1~TeV, solely by the total (isotropic 
equivalent) energy constraint of $\sim 10^{53}$ ergs, and by assuming the burst 
originated nearby ($z \ll 1$).

As was the case for GRB~941017, time-resolved spectral analyses of 
high-energy GRBs are needed in order to explicitly identify similar
distinct spectral component.  We have indeed performed such time-resolved 
analyses for all GRBs presented here, the result of which is the subject of
another paper that will follow this work (M.M.~Gonz\'alez et al. in preparation).
It is possible that the time-resolved spectral analysis of broadband spectra
would reveal such high-energy spectral components in many more GRBs.
Also in the MeV energy band and above, there should be a spectral cutoff 
due to pair-production attenuation \citep{bar00}.
Such a cutoff energy would determine the bulk Lorentz factor $\Gamma$, which is
one of the main factors, along with a redshift, preventing the 
determination of the intrinsic \peakenergy~values in the rest frame of 
expanding matter \citep{zham04}. 

Unfortunately, the TASC data are only available for the brightest of BATSE GRBs 
due to its sensitivity limitation.
Although GRB spectra with a distinct MeV component seem to be rare, and we
observed no pair-production cutoff in our analysis, these high-energy spectral
features could be sought after using currently existing very high-energy 
telescopes, such as AGILE\footnote{http://agile.iasf-roma.inaf.it/}, 
MAGIC\footnote{http://wwwmagic.mppmu.mpg.de/}, 
VERITAS\footnote{http://veritas.sao.arizona.edu/}, 
and HESS\footnote{http://www.mpi-hd.mpg.de/hfm/HESS/HESS.html}.
Upcoming observations by {\it GLAST}\footnote{http://glast.gsfc.nasa.gov/} with
much higher sensitivity in an unprecedented broad energy band of 
10~keV$-\sim$30~GeV, are anticipated to reveal broadband spectral 
characteristics of GRBs, which holds significant clues to understanding GRB
emission mechanism.



\newpage
%
\newpage
%

\newpage
\begin{figure}
\epsscale{1.1}
\plottwo{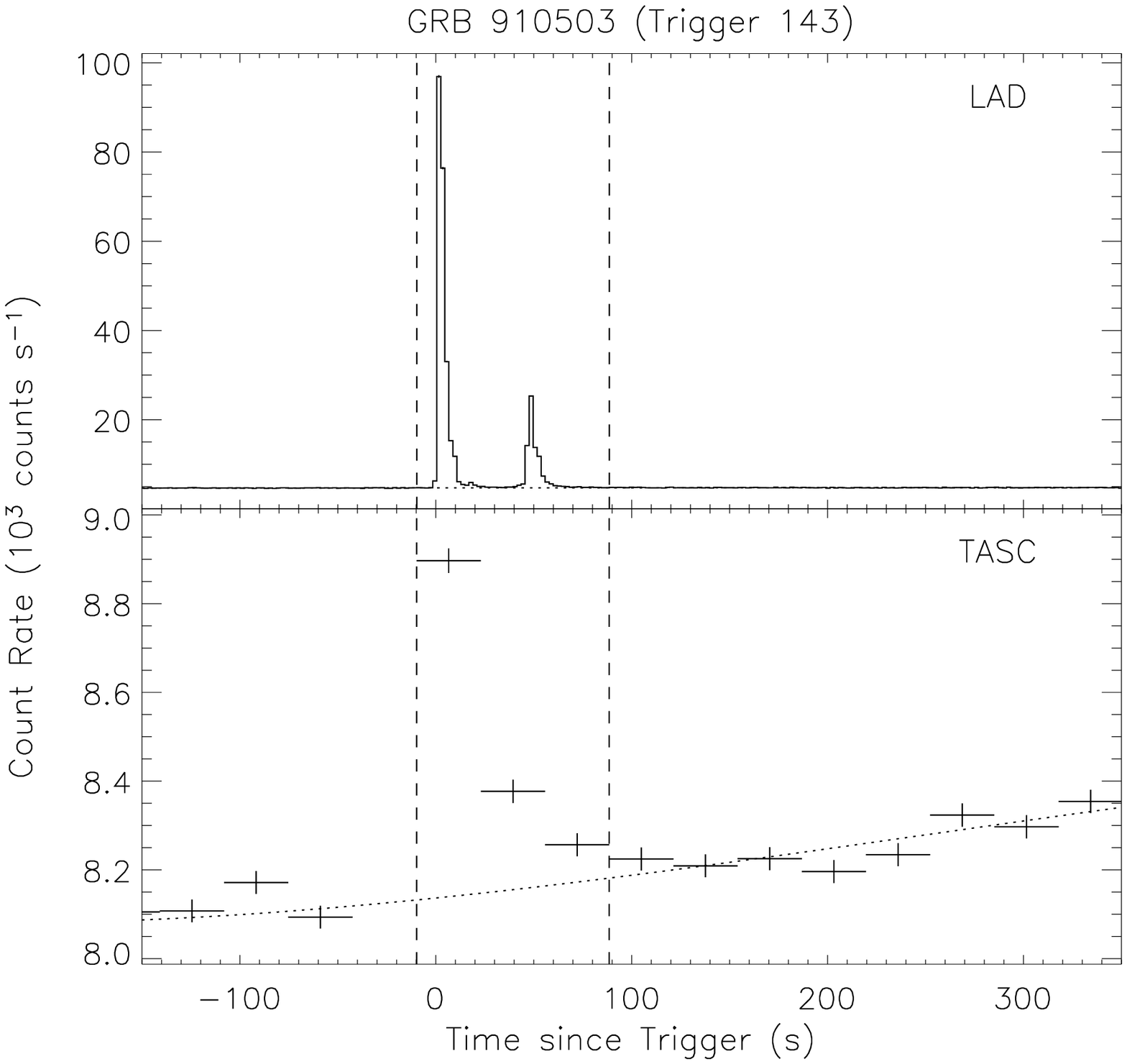}{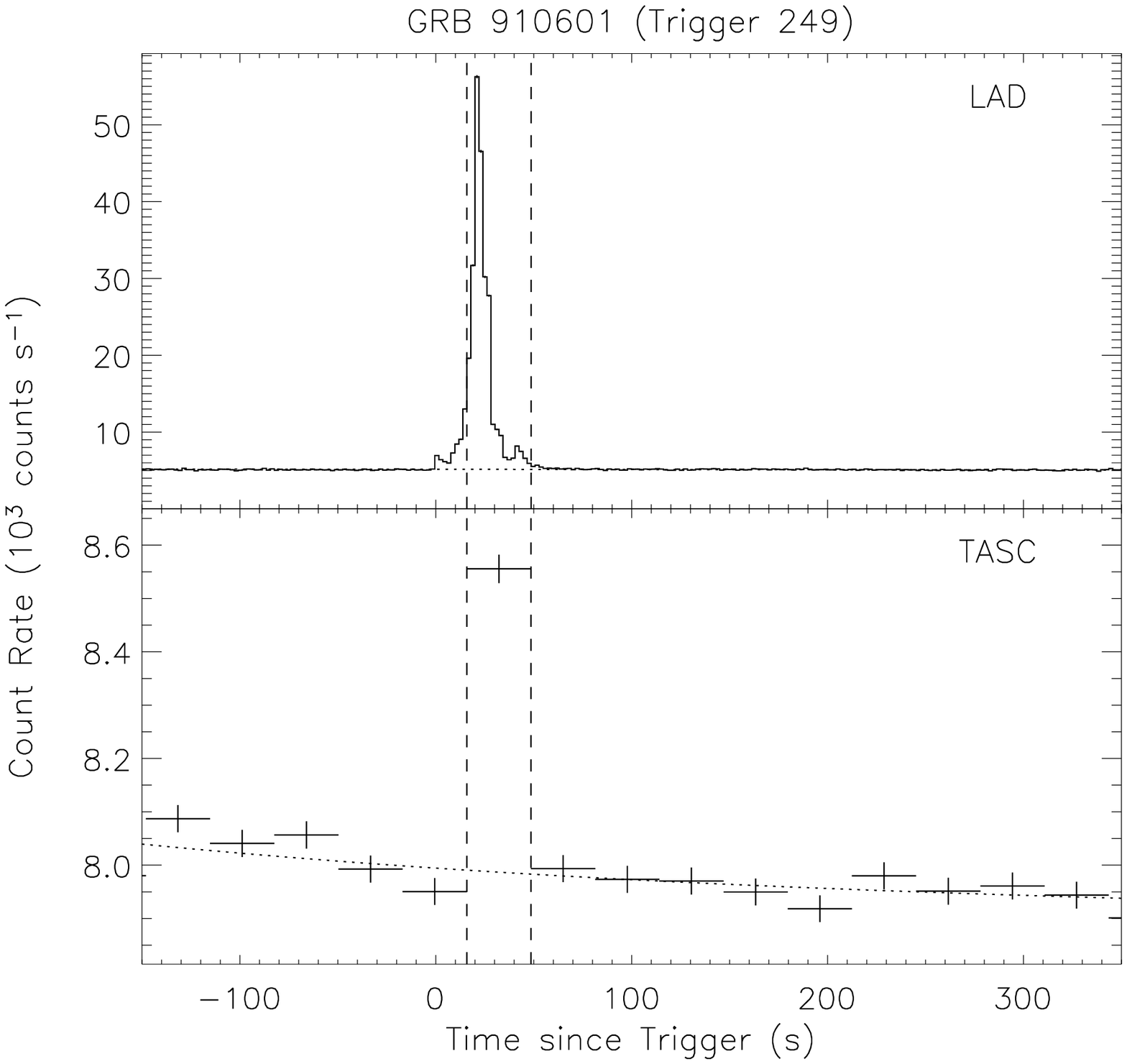}
\plottwo{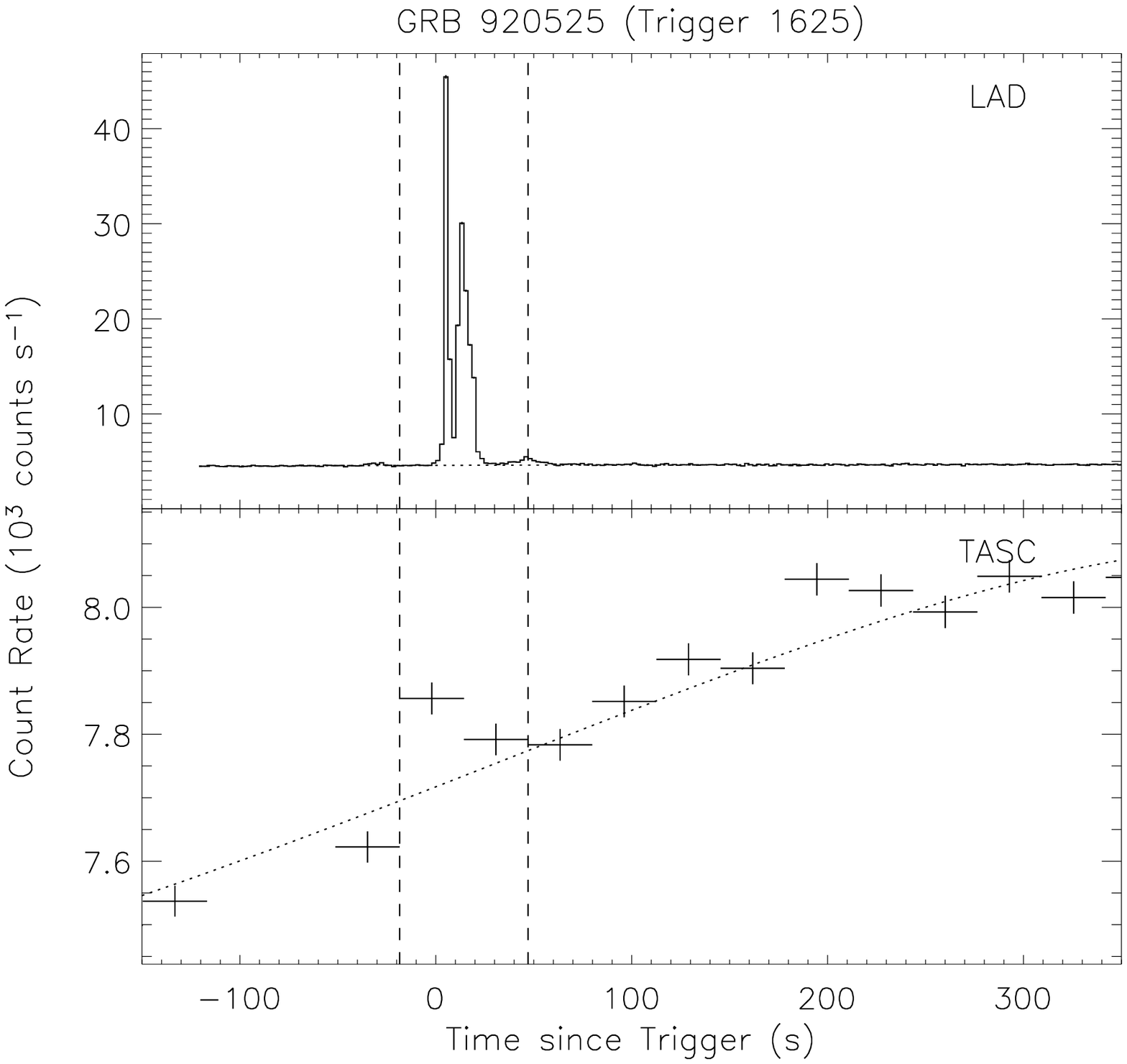}{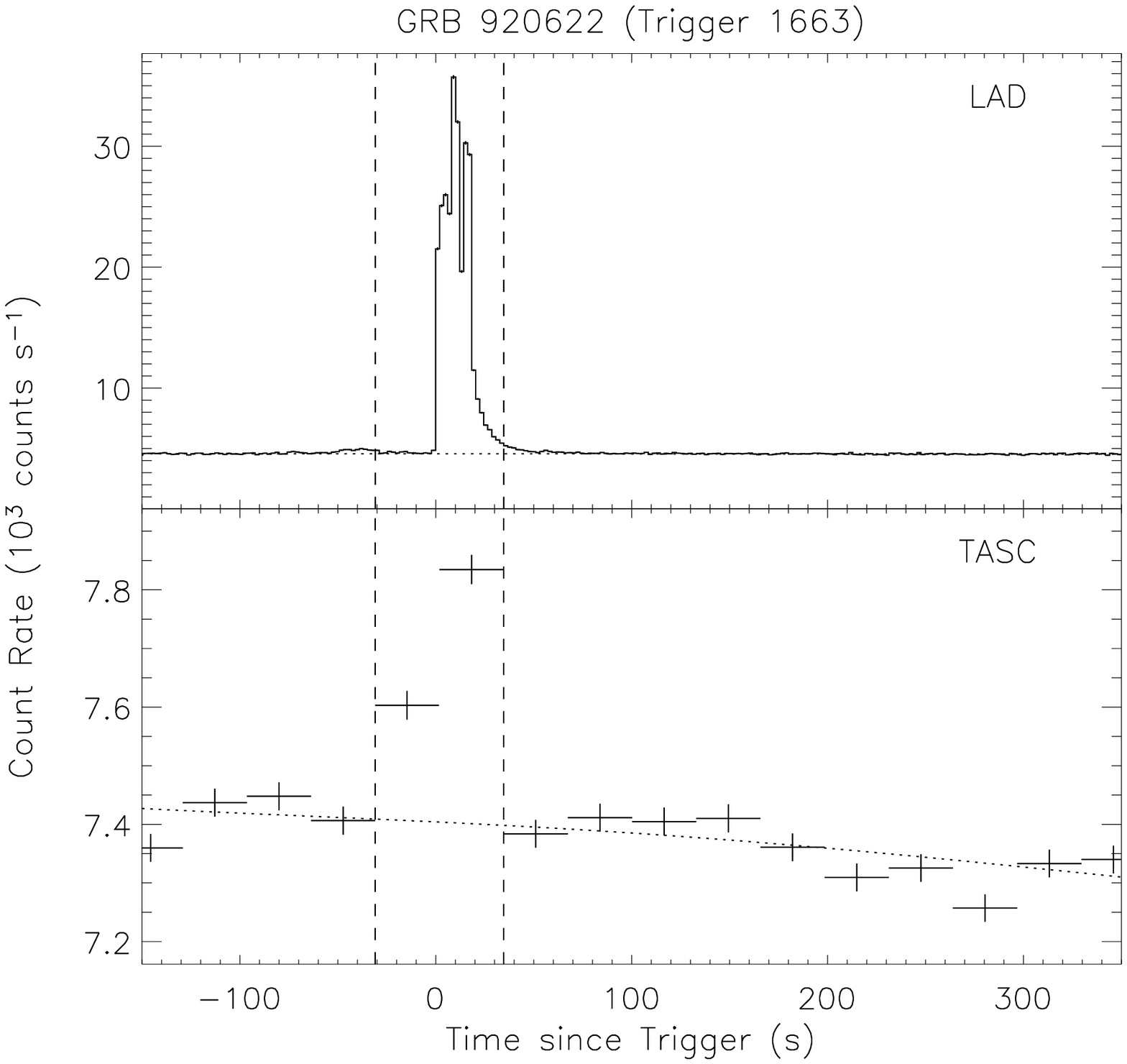}
\plottwo{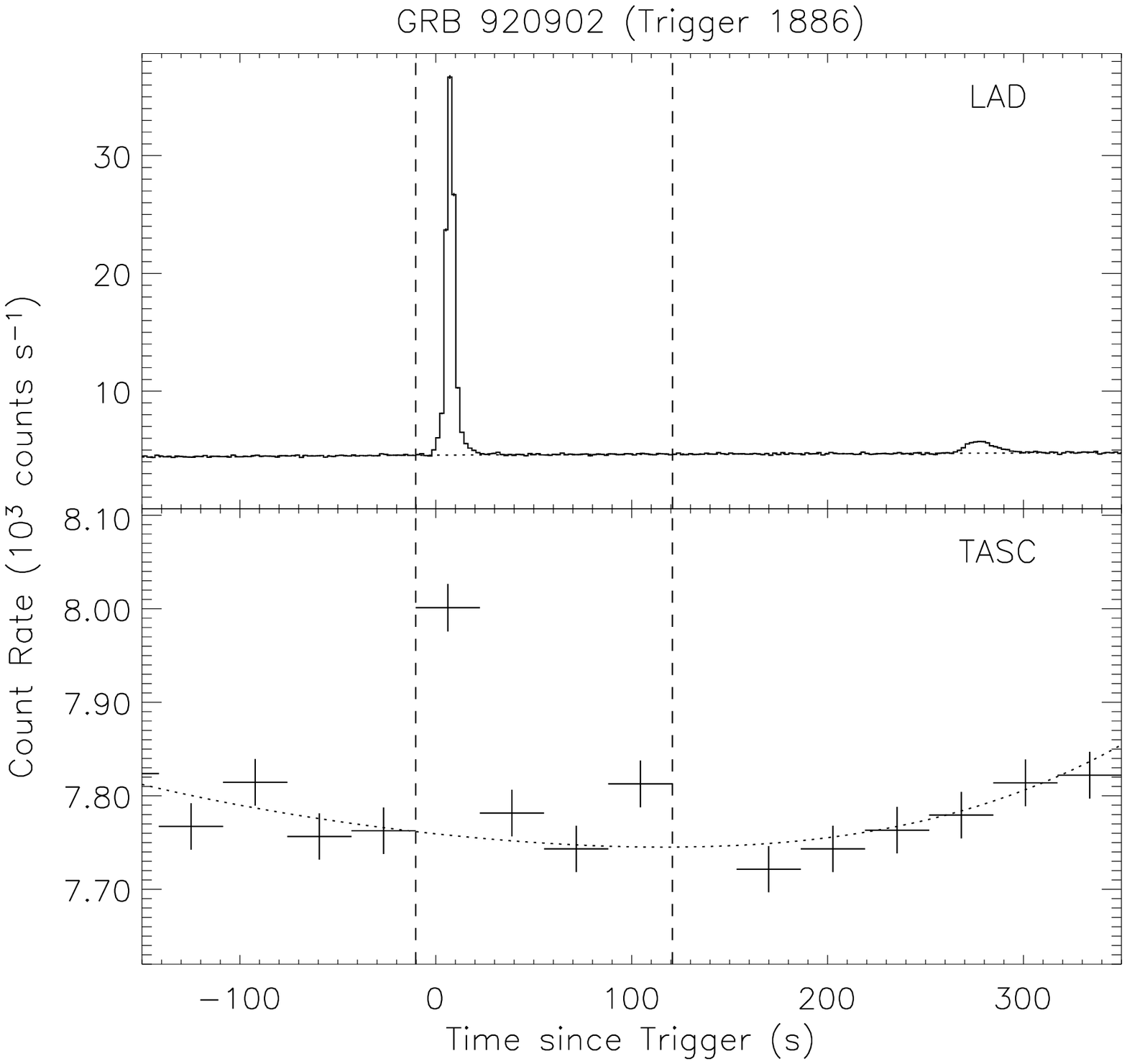}{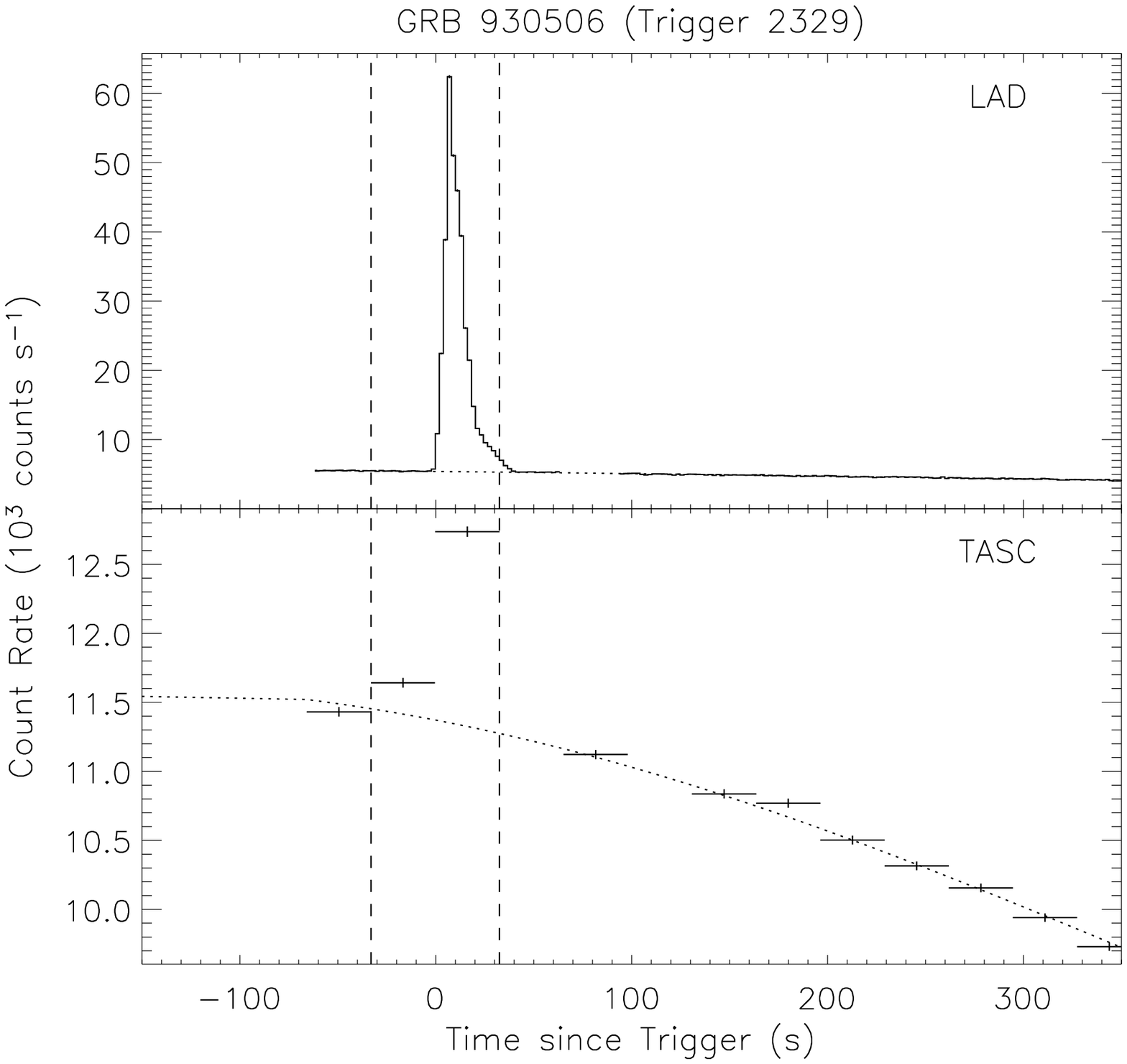}
\caption{LAD and TASC lightcurves of all 15 GRBs.
The dotted lines show background models and the vertical dashed lines indicate
the selected integration time intervals.}
\label{fig:lad_tasc_lc}
\end{figure}

%
\newpage
\begin{figure}
\figurenum{2}
\epsscale{1.1}
\plottwo{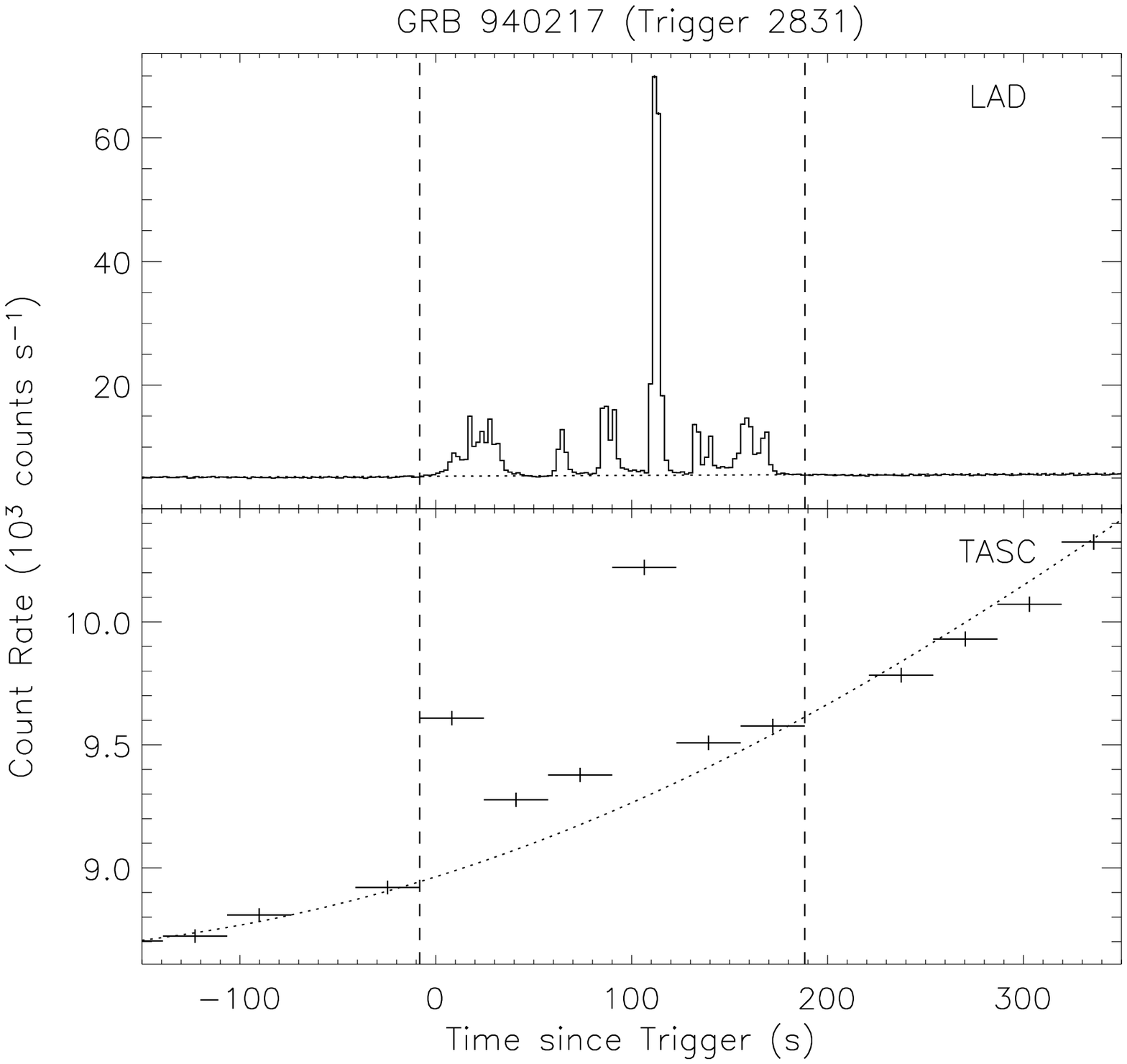}{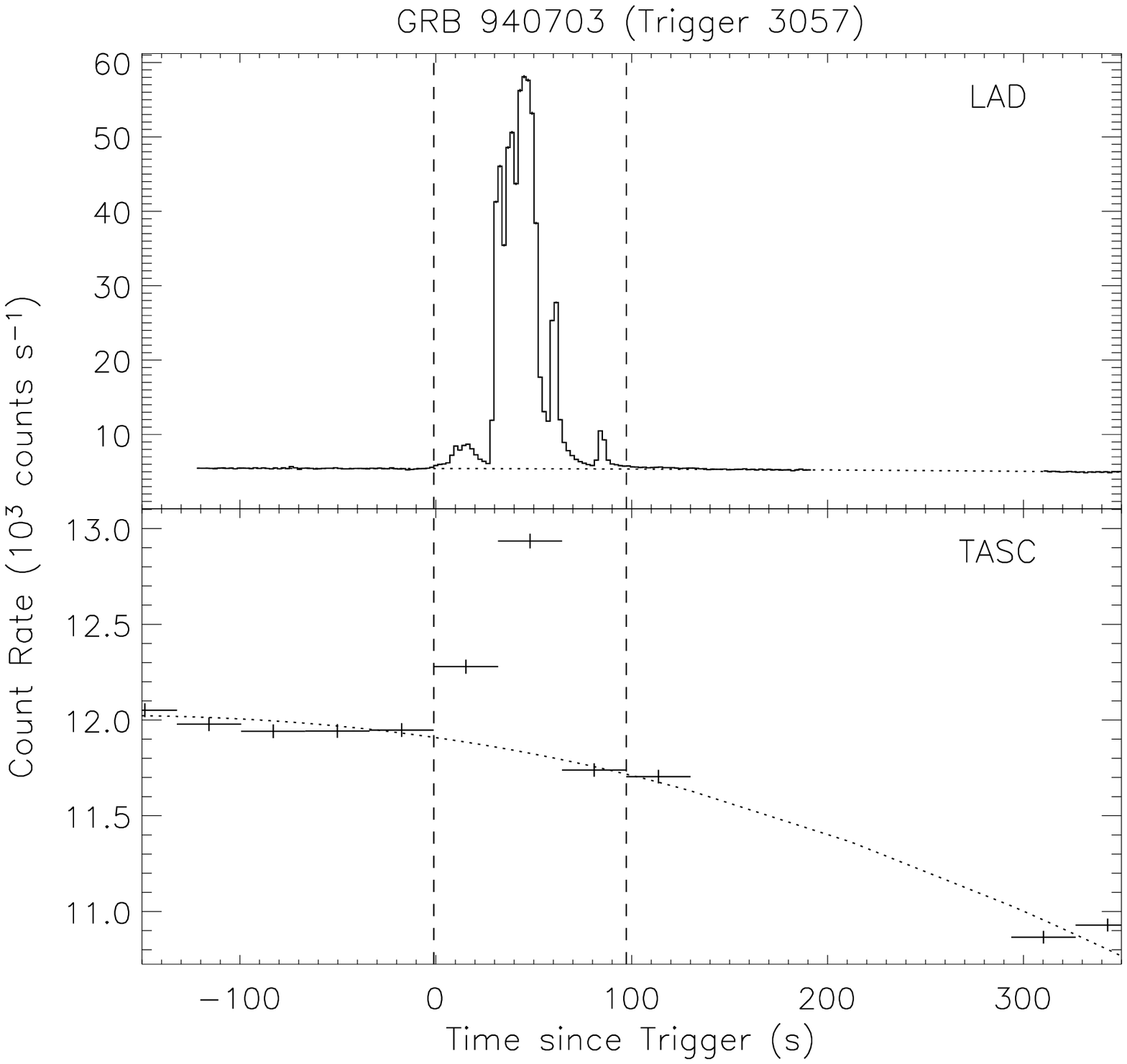}
\plottwo{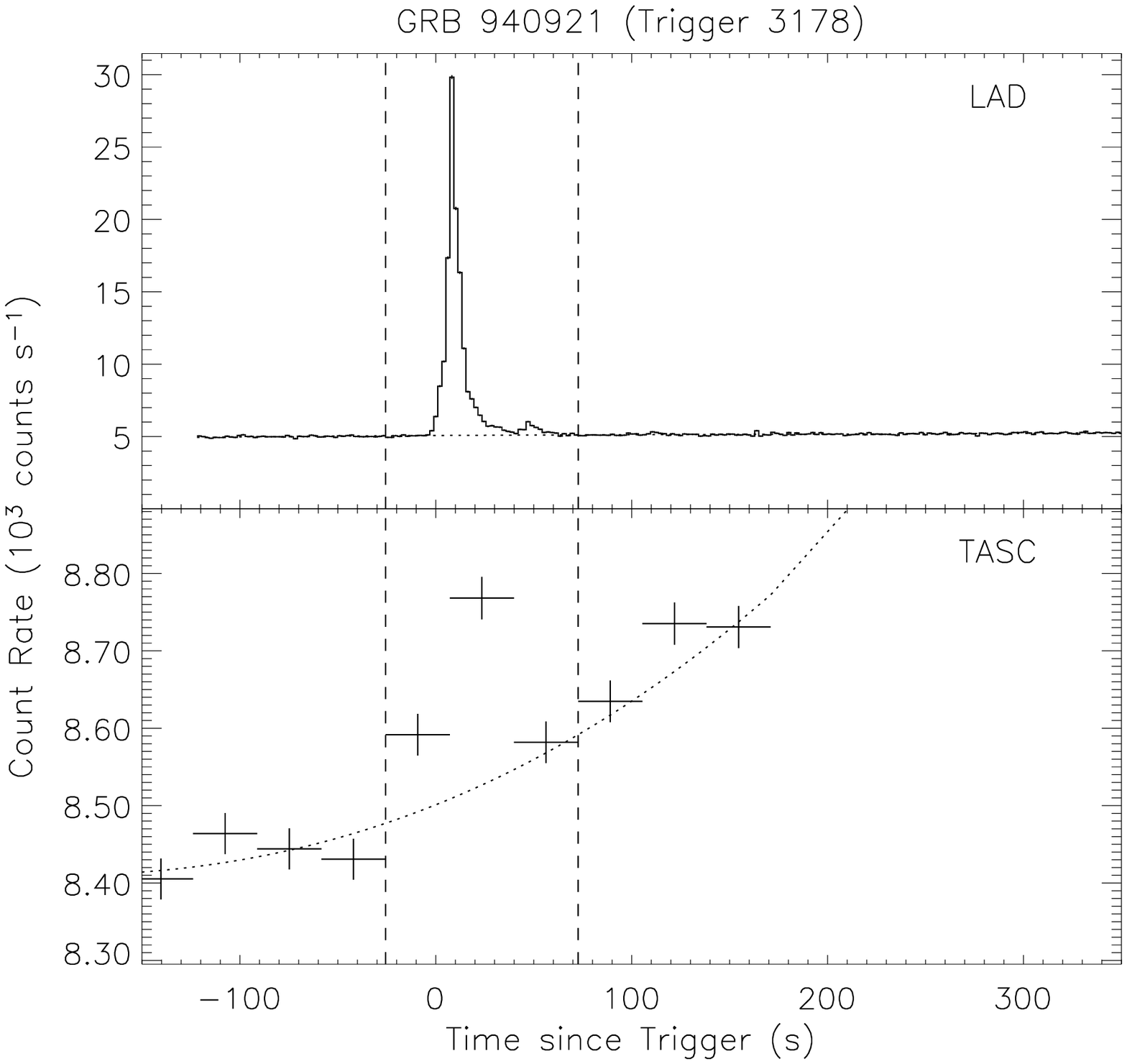}{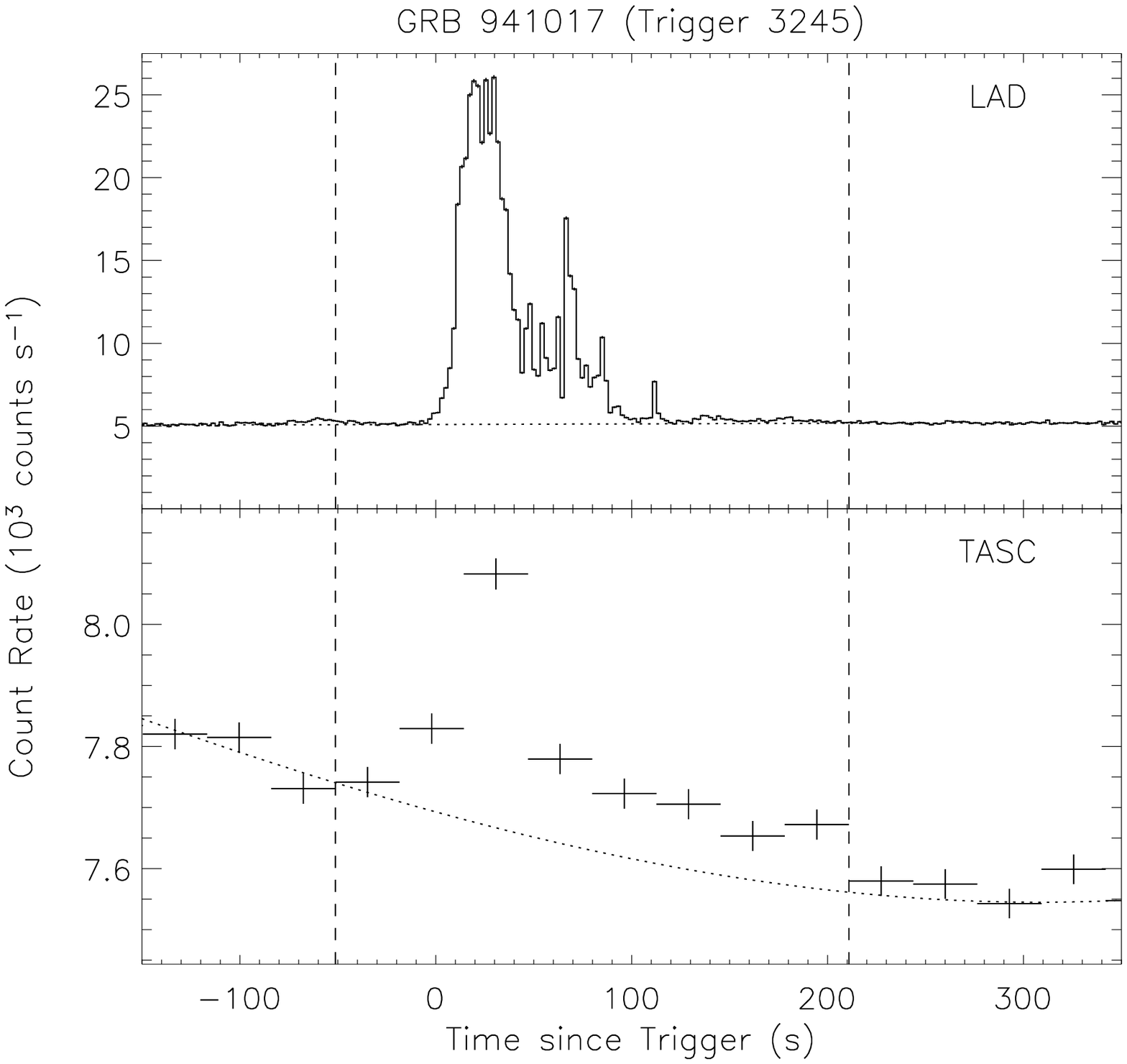}
\plottwo{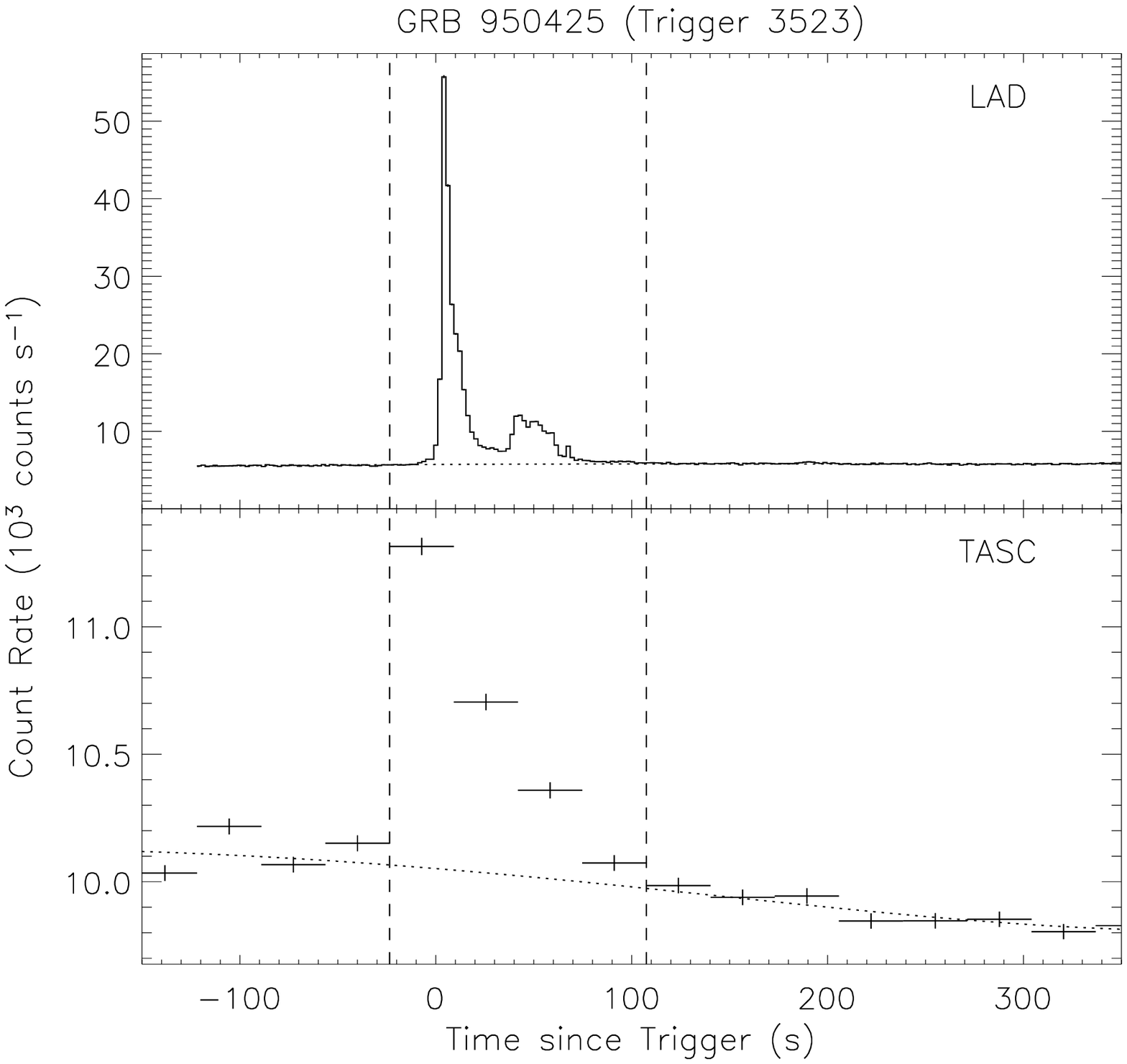}{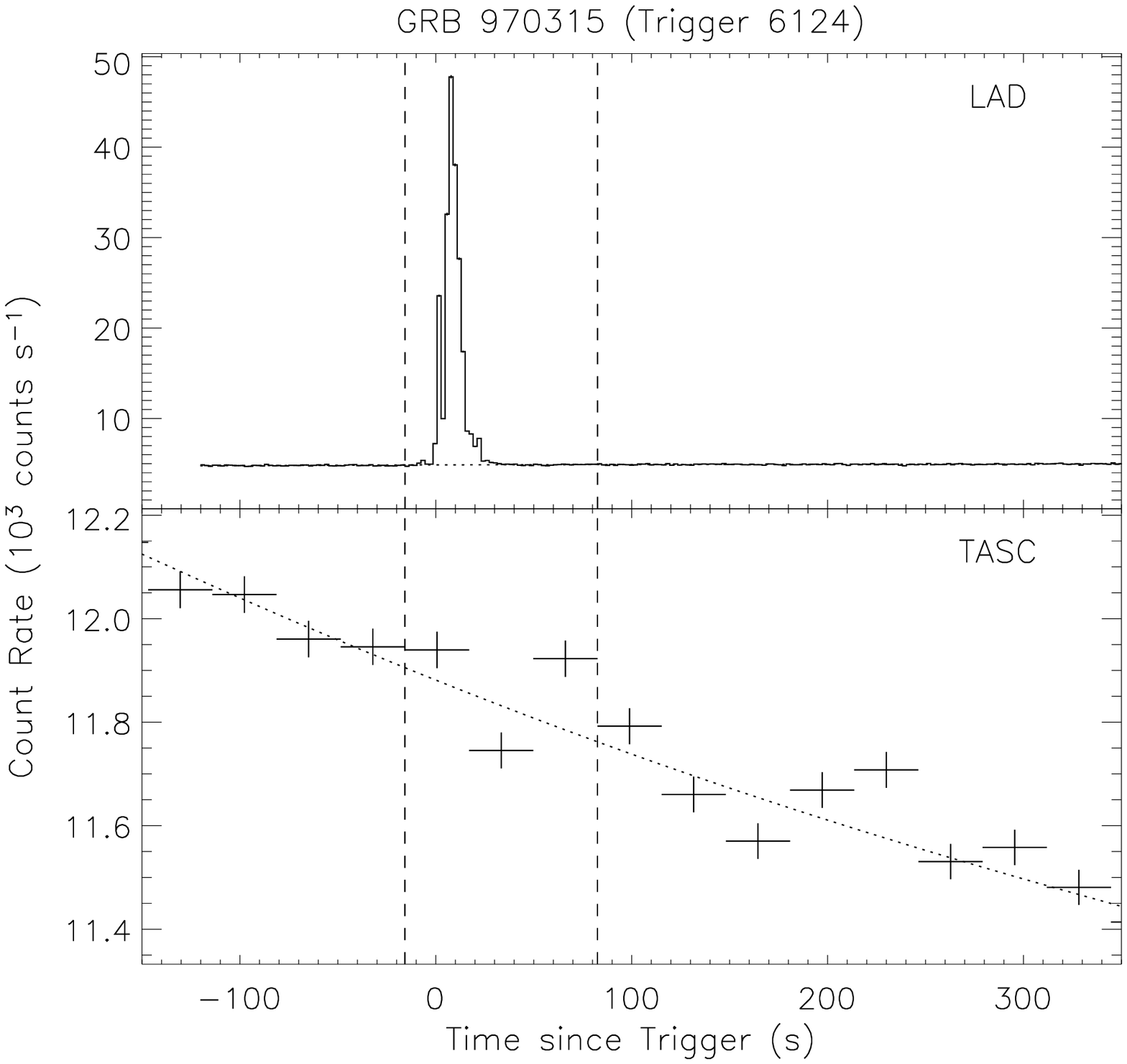}
\caption{Continued.}
\end{figure}
%
%

\newpage
\begin{figure}
\figurenum{2}
\epsscale{1.1}
\plottwo{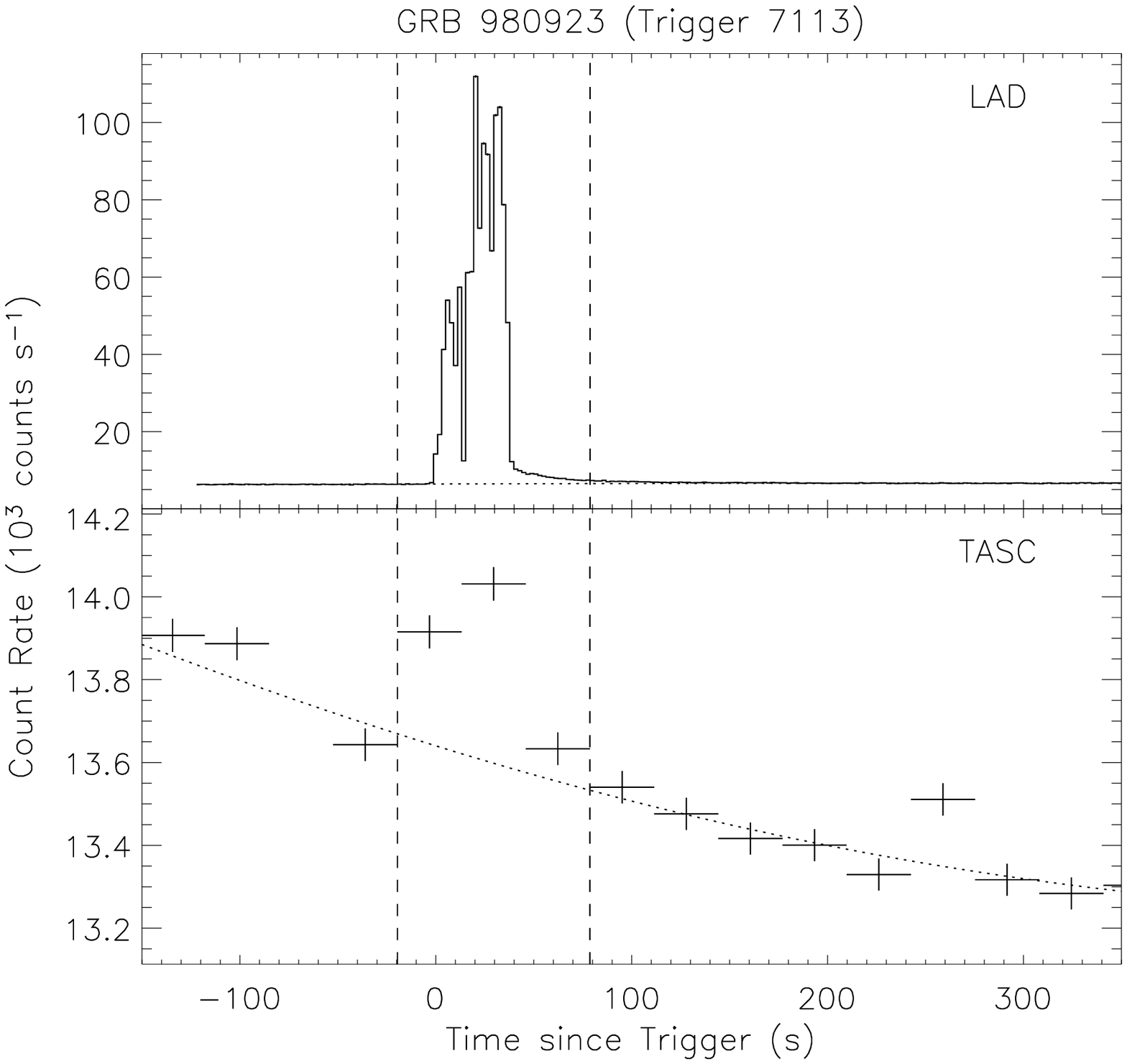}{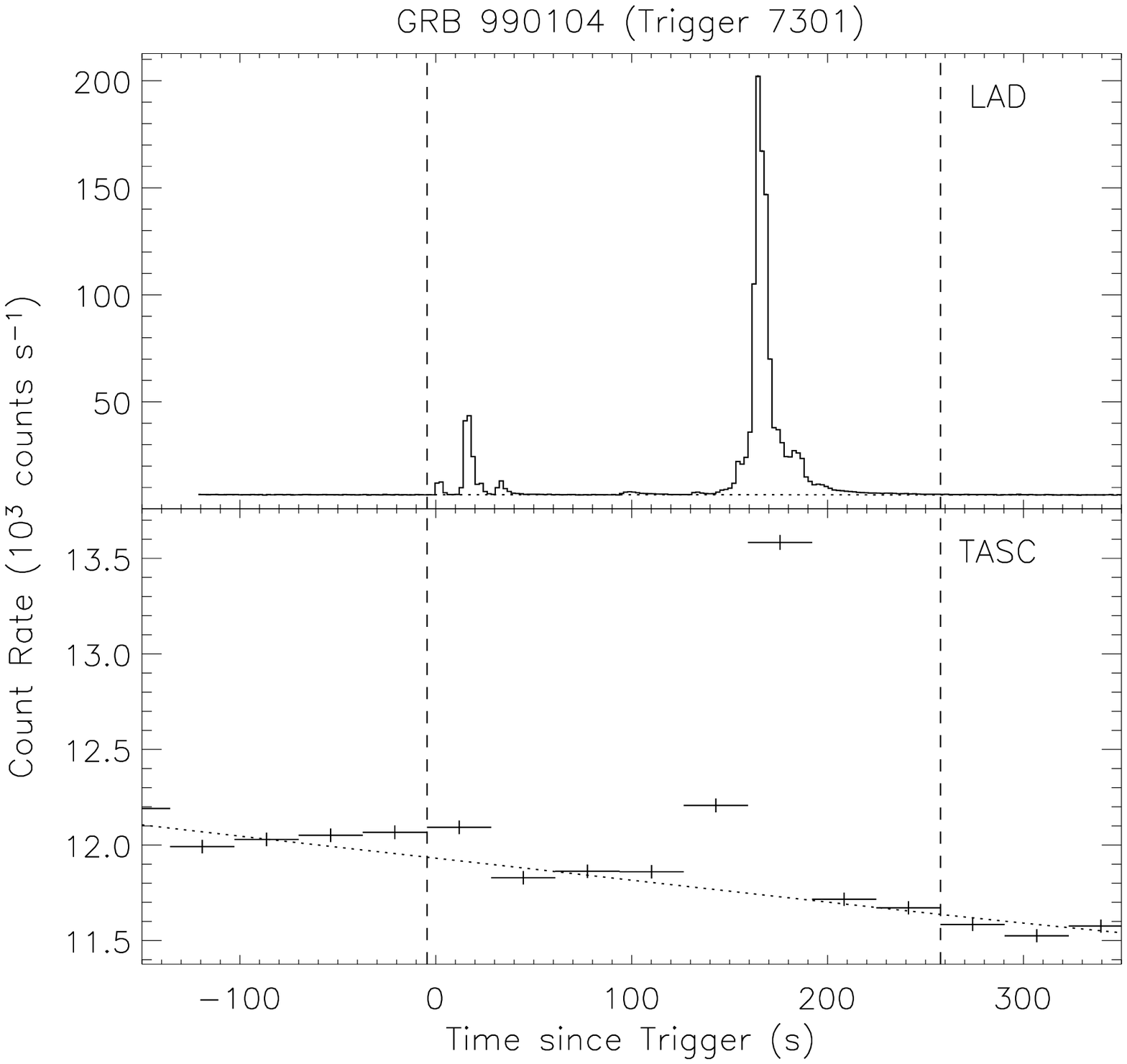} \\
\epsscale{0.5}
\plotone{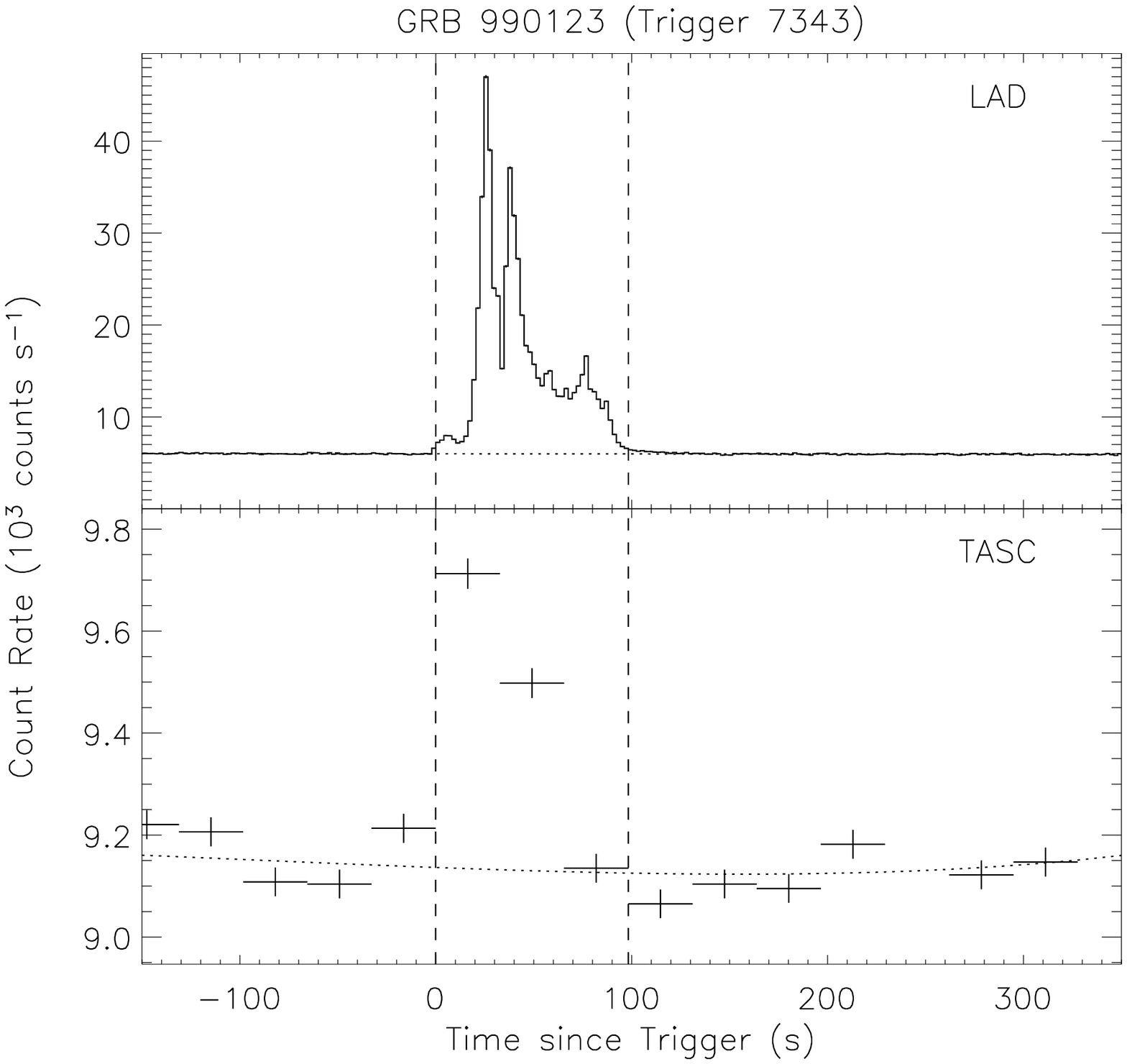}
\caption{Continued.}
\end{figure}
%

\newpage
\begin{figure}
\epsscale{0.97}
\plotone{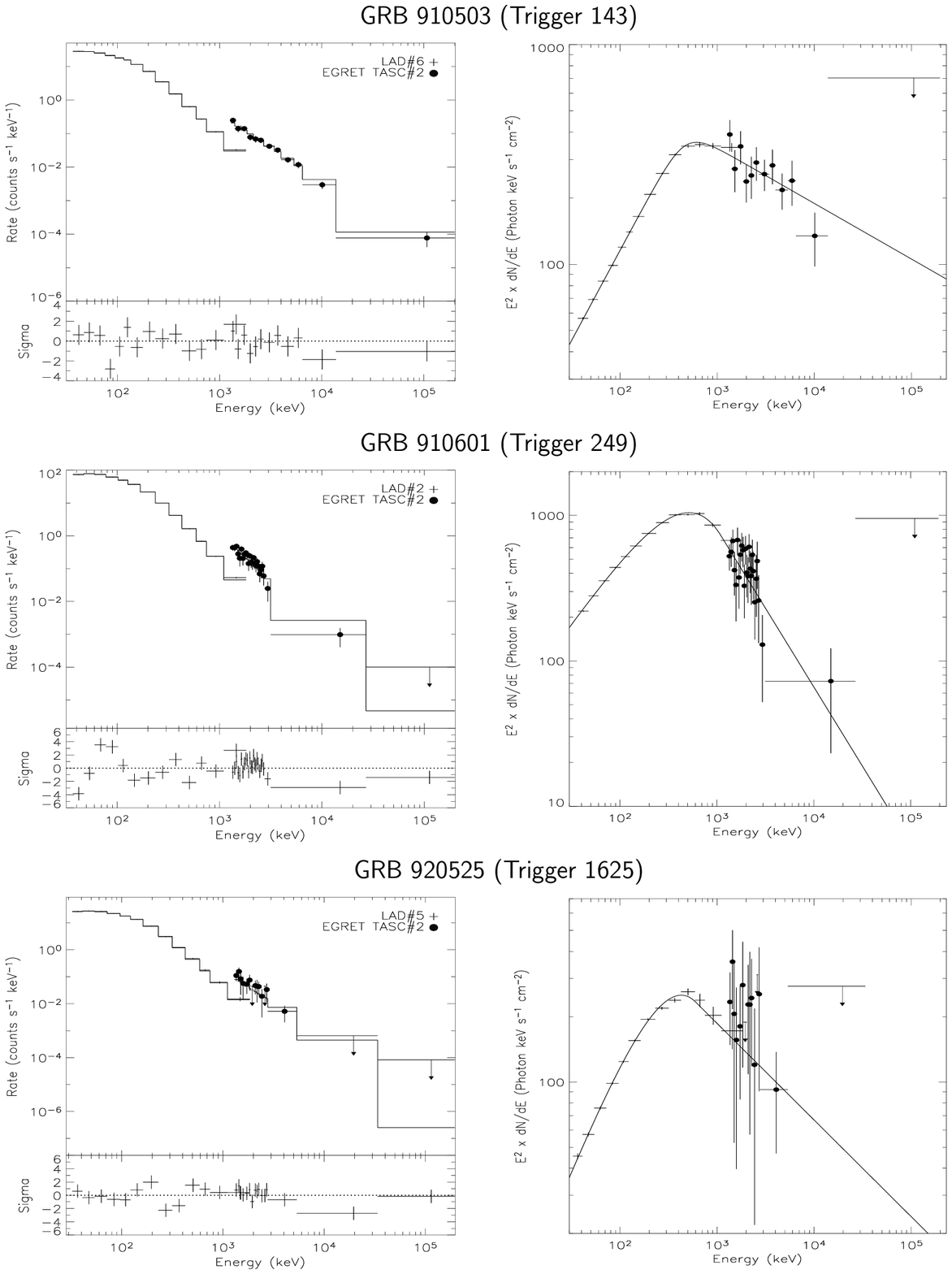}
\caption{Count spectra and the BEST models with sigma residuals 
({\it left panels}) and corresponding $\nu F_{\nu}$ spectra 
({\it right panels}) of LAD-TASC joint fits 
of all 15 GRBs. The BEST models are shown in solid lines in both panels.
The TASC data have been binned for display purpose only.}
\label{fig:lad_tasc_cntsp}
\end{figure}
%

\newpage
\begin{figure}
\figurenum{3}
\epsscale{0.97}
\plotone{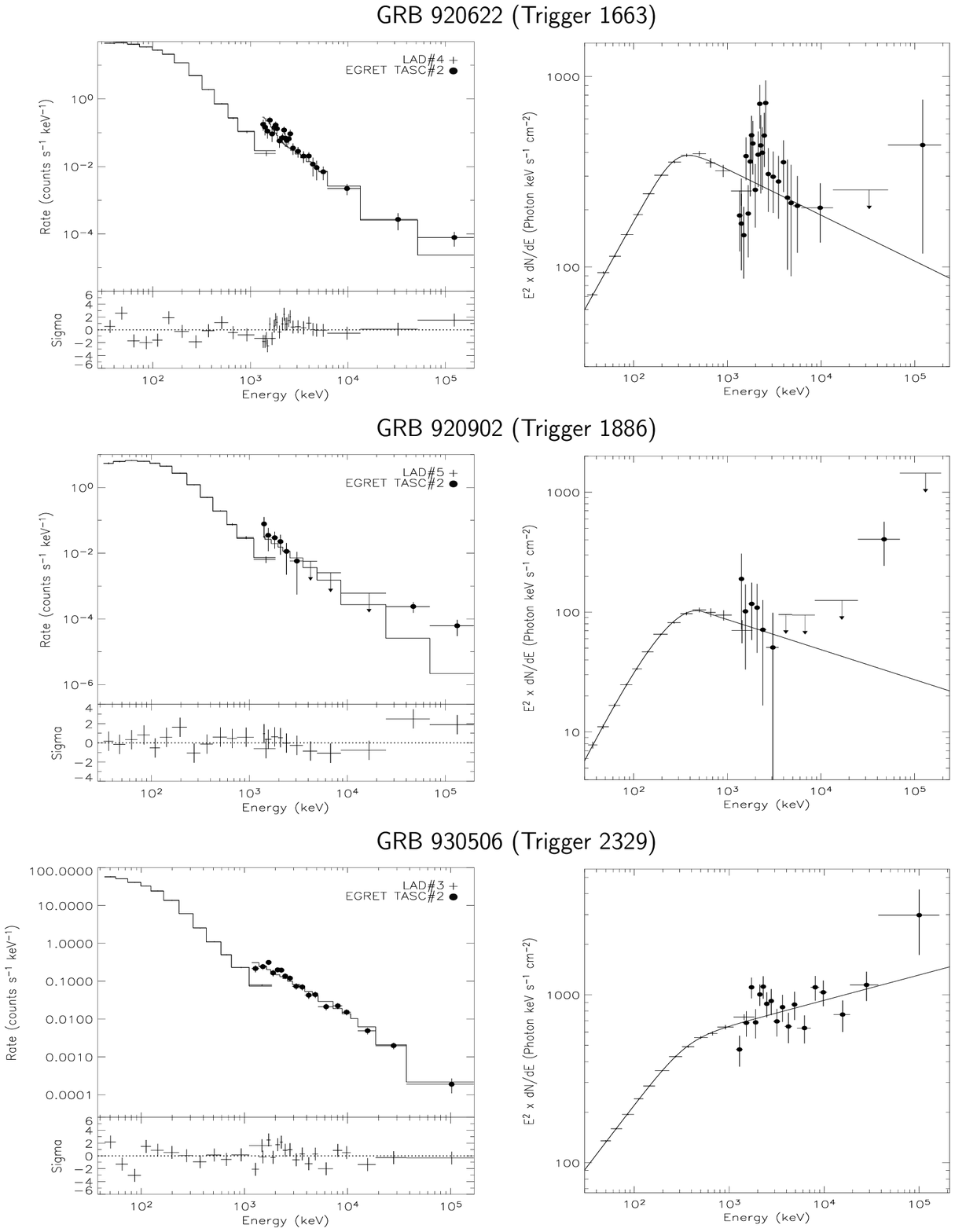}
\caption{Continued.}
\end{figure}
%

\newpage
\begin{figure}
\figurenum{3}
\epsscale{0.97}
\plotone{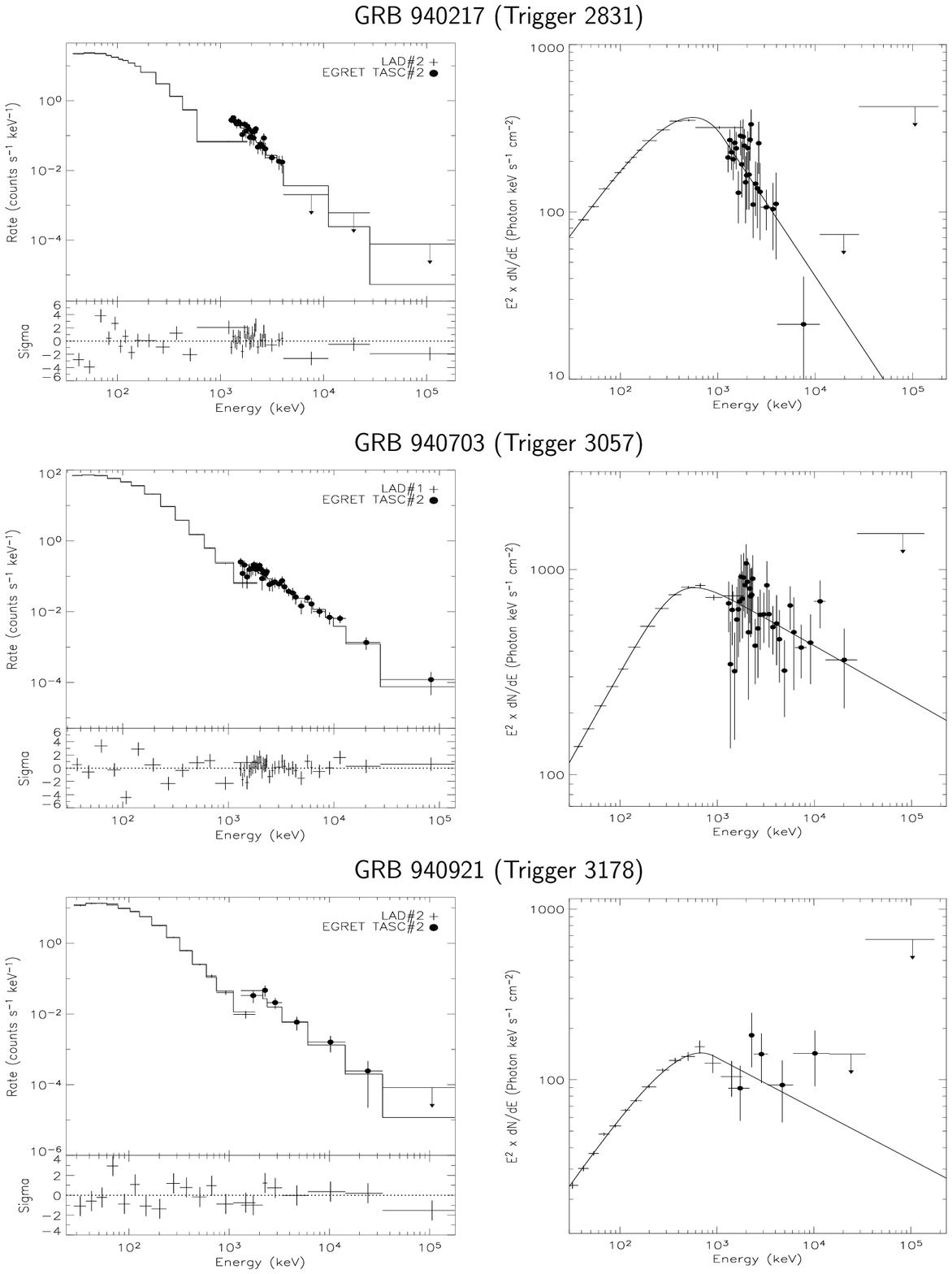}
\caption{Continued.}
\end{figure}

\newpage
\begin{figure}
\figurenum{3}
\epsscale{0.97}
\plotone{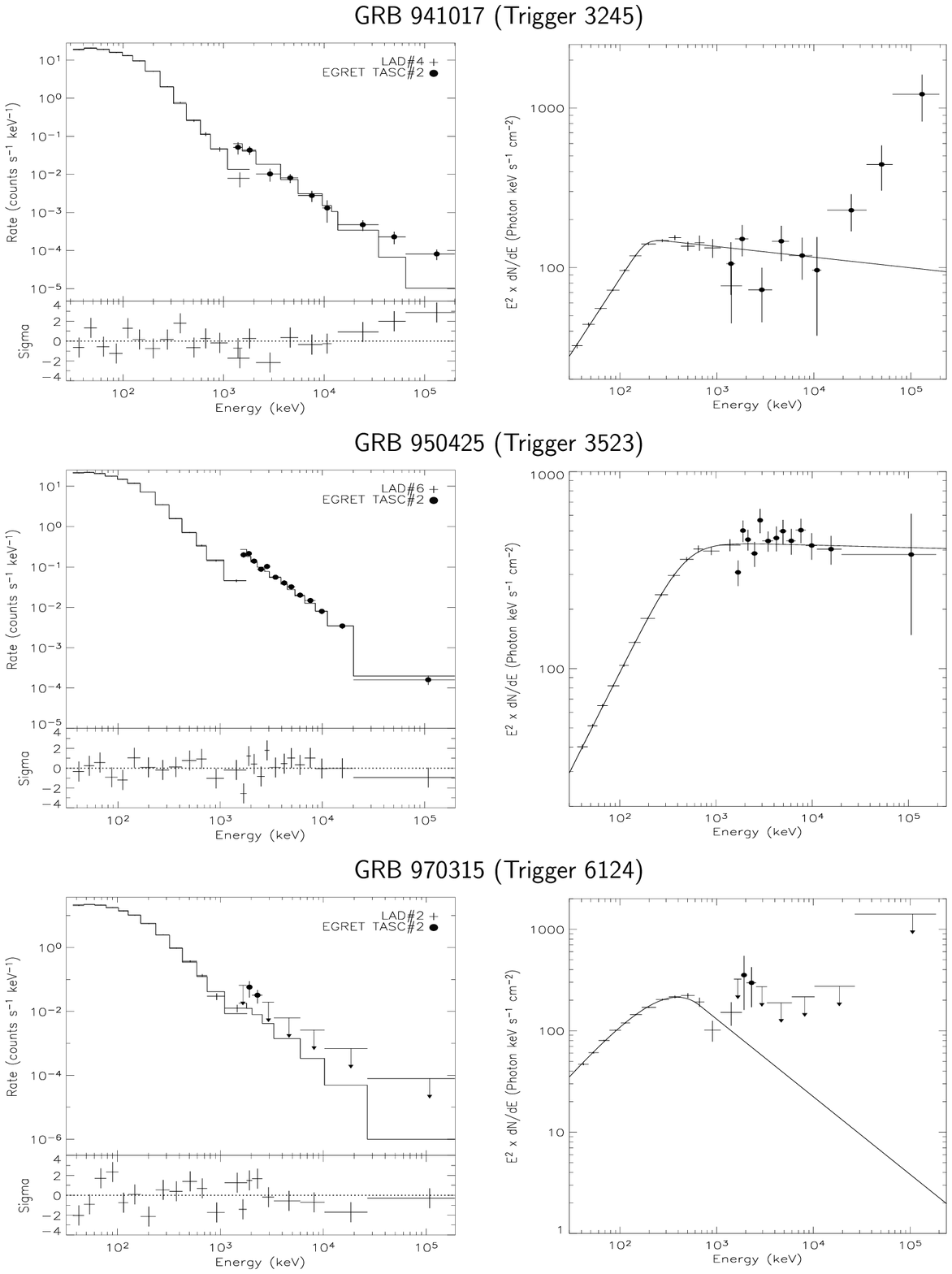}
\caption{Continued.}
\end{figure}

\newpage
\begin{figure}
\figurenum{3}
\epsscale{0.97}
\plotone{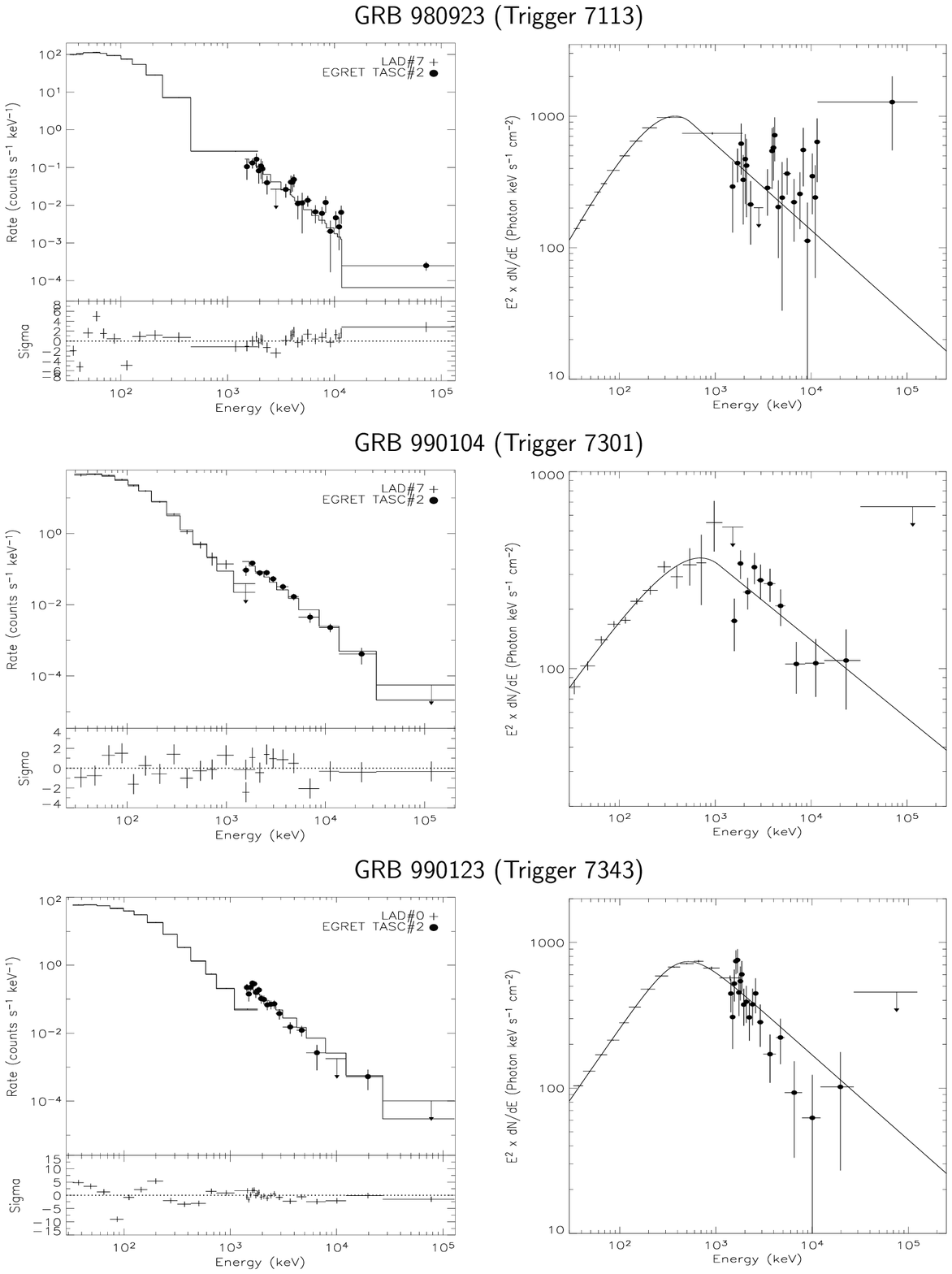}
\caption{Continued.}
\end{figure}

\newpage

\newpage

\newpage
\newpage
%

%
\begin{table}
\caption{Time and energy intervals used for the LAD--TASC joint spectral
         analysis of 15 GRBs.
         Time intervals are shown in units of second since BATSE trigger.} 
\label{tab:joint_events}
\tabcolsep=2pt
\begin{center}
\begin{tabular}{ccccccccccc}
\hline \hline \\[-2ex]
 & & & 
\multicolumn{2}{c}{Time Interval}   &&
\multicolumn{2}{c}{LAD Energy} && \multicolumn{2}{c}{TASC Energy}  \\
\cline{4-5}\cline{7-8}\cline{10-11}
 GRB & BATSE &  LAD &
 Start & End   && Start & End && Start & End \\
Name & Trigger No. & No. &
(s) & (s) && (keV) & (MeV) && (MeV) & (MeV) 
 \\[0.5ex] \hline
 \\[-1.8ex]
 910503 & ~143 & 6 & ~--9.8 & ~88.5 && 36.6 & 1.81 && 1.29 & 201 \\
 910601 & ~249 & 2 & ~~15.9 & ~48.6 && 37.0 & 1.82 && 1.32 & 198 \\
 920525 & 1625 & 5 & --18.4 & ~47.1 && 32.6 & 1.85 && 1.30 & 195 \\
 920622 & 1663 & 4 & --30.9 & ~34.6 && 31.9 & 1.81 && 1.32 & 195 \\
 920902 & 1886 & 5 & --10.2 & 120.8 && 32.7 & 1.85 && 1.39 & 196 \\
 930506 & 2329 & 3 & --33.1 & ~32.5 && 44.2 & 1.85 && 1.18 & 167 \\
 940217 & 2831 & 2 & ~--8.2 & 188.4 && 37.1 & 1.82 && 1.25 & 187 \\
 940703 & 3057 & 1 & ~--1.0 & ~97.2 && 33.3 & 1.88 && 1.28 & 139 \\
 940921 & 3178 & 2 & --25.6 & ~72.6 && 28.1 & 1.82 && 1.31 & 178 \\
 941017 & 3245 & 4 & --51.2 & 210.9 && 31.8 & 1.81 && 1.25 & 199 \\
 950425 & 3523 & 6 & --23.6 & 107.5 && 36.3 & 1.80 && 1.57 & 197 \\
 970315 & 6124 & 2 & --15.8 & ~82.5 && 37.1 & 1.82 && 1.51 & 189 \\
 980923 & 7113 & 7 & --20.0 & ~78.7 && 33.4 & 1.92 && 1.44 & 131 \\
 990104 & 7301 & 7 & ~--4.5 & 257.7 && 29.3 & 1.96 && 1.45 & 202 \\
 990123 & 7343 & 0 & ~--0.1 & ~98.3 && 33.1 & 1.81 && 1.40 & 128 \\
\hline
\end{tabular}
\end{center}
\end{table}
%
%
\begin{table}
\caption{Summary of LAD--TASC joint fit results for time-integrated spectra.
         Full functional form and parameter description of each spectral model 
         can be found in \kan. 
         1$\sigma$ uncertainties are shown in parentheses.} 
\label{tab:joint_results}
\scriptsize
\begin{center}
\begin{tabular}{ccccccccccc}
\hline \hline \\[-2ex]
    &     &
 \multicolumn{9}{c}{Spectral Fit Parameters} \\ \cline{4-11}\\[-2ex]
 GRB & \hspace{-3ex}BATSE\hspace{-3ex} & BEST & Amplitude & \peakenergy$^{^1}$
 & Low Index$^{^2}$ 
 & High Index$^{^3}$ &  $E_{\rm b}^{^4}$
 & \hspace{-3ex}Break \hspace{-3ex} & EAC & $\chi^2$/dof 
 \\
 Name & \hspace{-3ex}Trig No.\hspace{-3ex} & Model & (ph~s$^{-1}$~cm$^{-2}$) 
 & (keV) 
 & 
 &  &  (keV) 
 & \hspace{-3ex}Scale, $\Lambda$ \hspace{-3ex} &  &
 \\[0.5ex] \hline
 \\[-1.8ex]
 910503 & ~143 & SBPL & 0.0116 (0.0000) & 612 ( 28) & --1.18 (0.01) & --2.25 (0.03) & 467 ( 17) & 0.20 & 0.63 & 204.4/224 \\ 
 910601 & ~249 & BAND & 0.0567 (0.0003) & 515 ( ~8) & --1.04 (0.01) & --3.08 (0.10) & 562 ( 26) & ---  & 0.71 & 288.0/221 \\ 
 920525 & 1625 & BAND & 0.0151 (0.0003) & 427 ( 18) & --0.88 (0.03) & --2.45 (0.10) & 311 ( 23) & ---  & 0.19 & 250.1/221 \\ 
 920622 & 1663 & SBPL & 0.0174 (0.0001) & 386 ( 20) & --1.11 (0.01) & --2.24 (0.04) & 286 ( 11) & 0.20 & 0.73 & 228.9/220 \\ 
 920902 & 1886 & BAND & 0.0043 (0.0001) & 469 ( 26) & --0.41 (0.05) & --2.25 (0.11) & 284 ( 21) & ---  & 0.70 & 110.8/219 \\ 
 930506 & 2329 & SBPL & 0.0220 (0.0001) & $>$167000?& --1.26 (0.01) & --1.85 (0.02) & 349 ( 22) & 0.30 & 0.50 & 226.6/208 \\ 
 940217 & 2831 & BAND & 0.0203 (0.0001) & 553 ( 12) & --1.17 (0.01) & --2.88 (0.10) & 580 ( 35) & ---  & 0.49 & 257.7/222 \\ 
 940703 & 3057 & SBPL & 0.0315 (0.0001) & 585 ( 29) & --1.15 (0.01) & --2.27 (0.03) & 393 ( 13) & 0.30 & 0.33 & 211.8/206 \\ 
 940921 & 3178 & BAND & 0.0067 (0.0002) & 684 ( 65) & --1.17 (0.03) & --2.30 (0.13) & 478 ( 74) & ---  & 0.73 & 218.6/217 \\
 941017 & 3245 & SBPL & 0.0086 (0.0001) & 250 ( 17) & --1.07 (0.02) & --2.07 (0.03) & 185 ( ~7) & 0.10 & 0.44 & 183.4/215 \\ 
 950425 & 3523 & SBPL & 0.0095 (0.0001) &2150 (1568)& --1.03 (0.02) & --2.01 (0.03) & 479 ( 30) & 0.30 & 0.69 & 203.8/215 \\ 
 970315 & 6124 & BAND & 0.0143 (0.0004) & 392 ( 19) & --0.90 (0.03) & --2.77 (0.29) & 345 ( 55) & ---  & 0.63 & 207.3/215 \\ 
 980923 & 7113 & BAND & 0.0639 (0.0005) & 379 ( ~5) & --0.66 (0.01) & --2.65 (0.05) & 294 ( ~8) & ---  & 0.37 & 267.7/198 \\ 
 990104 & 7301 & BAND & 0.0190 (0.0006) & 696 (119) & --1.31 (0.05) & --2.40 (0.09) & 560 (104) & ---  & 0.35 & 202.5/217 \\
 990123 & 7343 & SBPL & 0.0256 (0.0001) & 517 ( 16) & --1.04 (0.01) & --2.58 (0.03) & 436 ( 10) & 0.30 & 0.68 & 400.4/200 \\ 
\hline
\end{tabular}
\begin{minipage}{0.98\textwidth}
\small{$^1$ Energy at which the $\nu F_{\nu}$ spectrum peaks.
            For BAND, it is a {\it Fitted} \epeak, if $\beta < -2$, and
            for SBPL, it is a {\it calculated} \epeak~(see Appendix B in \kan).}\\
\small{$^2$ $\alpha$ for BAND and $\lambda_1$ for SBPL.}\\
\small{$^3$ $\beta$ for BAND and $\lambda_2$ for SBPL.}\\
\small{$^4$ {\it Fitted} \eb~for SBPL, and {\it calculated} \eb~ 
         for BAND (see Appendix C in \kan).}
\end{minipage}
\end{center}
\end{table}

\end{document}